\documentclass[11pt,a4paper]{article}
\usepackage{amsfonts}
\usepackage{amsmath,amssymb}
\usepackage{epsfig,graphicx}

\input{tcilatex}

\begin{document}

\begin{center}
{\LARGE \ Spontaneously Generated Tensor Field Gravity}

\bigskip

\bigskip

\bigskip

\textbf{J.L.~Chkareuli}$^{1,2}$\textbf{, C.D. Froggatt}$^{3}$\textbf{, H.B.
Nielsen}$^{4}$\bigskip

$^{1}$\textit{Center for Elementary Particle Physics, ITP, Ilia State
University, 0162 Tbilisi, Georgia}

$^{2}$\textit{Particle Physics Department, \textit{A}ndronikashvili
Institute of Physics, 0177 Tbilisi, Georgia\ }

$^{3}$\textit{Department of Physics and Astronomy, Glasgow University,
Glasgow G12 8QQ, Scotland\vspace{0pt}\\[0pt]
}

$^{4}$\textit{Niels Bohr Institute, Blegdamsvej 17-21, DK 2100 Copenhagen,
Denmark}

\bigskip

\bigskip

\bigskip \bigskip

\bigskip \textbf{Abstract}

\bigskip
\end{center}

An arbitrary local theory of a symmetric two-tensor field $H_{\mu \nu }$ in
Minkowski spacetime is considered, in which the equations of motion are
required to be compatible with a nonlinear length-fixing constraint $H_{\mu
\nu }^{2}=\pm M^{2}$ leading to spontaneous Lorentz invariance violation,
SLIV ($M$ is the proposed scale for SLIV). Allowing the parameters in the
Lagrangian to be adjusted so as to be consistent with this constraint, the
theory turns out to correspond to linearized general relativity in the weak
field approximation, while some of the massless tensor Goldstone modes
appearing through SLIV are naturally collected in the physical graviton. In
essence the underlying diffeomophism invariance emerges as a necessary
condition for the tensor field $H_{\mu \nu }$ not to be superfluously
restricted in degrees of freedom, apart from the constraint due to which the
true vacuum in the theory is chosen by SLIV. The emergent theory appears
essentially nonlinear, when expressed in terms of the pure Goldstone tensor
modes and contains a plethora of new Lorentz and $CPT$ violating couplings.
However, these couplings do not lead to physical Lorentz violation once this
tensor field gravity is properly extended to conventional general relativity.

\bigskip

\bigskip



\thispagestyle{empty}\newpage

\section{Introduction}

It is conceivable that spontaneous Lorentz invariance violation (SLIV) could
provide a dynamical approach to quantum electrodynamics, gravity and
Yang-Mills theories with the photon, graviton and gluons appearing as
massless Nambu-Goldstone bosons\ \cite{bjorken,ph,eg,suz} (for some later
developments see \cite{cfn,kraus,kos,car,cjt})\footnote{%
Independently of the problem of the origin of local symmetries, Lorentz
violation in itself has attracted considerable attention as an interesting
phenomenological possibility which may be probed in direct Lorentz
non-invariant extensions of quantum electrodynamics (QED) and the Standard
Model \cite{chadha,jakiw,alan,glashow,ted}.}. However, in contrast to the
spontaneous violation of internal symmetries, SLIV seems not to necessarily
imply a physical breakdown of Lorentz invariance. Rather, when appearing in
a gauge theory framework, this may eventually result in a noncovariant gauge
choice in an otherwise gauge invariant and Lorentz invariant theory. In
substance the SLIV ansatz, due to which the vector field develops a vacuum
expectation value (vev) $<A_{\mu }(x)>$ $=n_{\mu }M$ \ (where $n_{\mu }$ is
a properly oriented unit Lorentz vector, while $M$ is the proposed SLIV
scale), may itself be treated as a pure gauge transformation with a gauge
function linear in coordinates, $\omega (x)=$ $n_{\mu }x^{\mu }M$. In this
sense, gauge invariance in QED leads to the conversion of SLIV into gauge
degrees of freedom of the massless Goldstonic photon, unless it is disturbed
by some extra (potential-like) terms. This is what one could refer to as the
generic non-observability of SLIV in QED. Moreover, as was shown some time
ago \cite{cfn}, gauge theories, both Abelian and non-Abelian, can be
obtained by themselves from the requirement of the physical
non-observability of SLIV induced by vector fields rather than from the
standard gauge principle.

A possible source for such a kind of unobserved SLIV is \textquotedblleft
nonlinearly realized\textquotedblright\ Lorentz symmetry imposed just by
postulate on an underlying vector field $A_{\mu }$ through the length-fixing
constraint 
\begin{equation}
A_{\mu }A^{\mu }=n^{2}M^{2}\text{ , \ \ }n^{2}\equiv n_{\nu }n^{\nu }=\pm 1,
\label{const}
\end{equation}%
rather than due to some vector field potential. This constraint was first
studied in the gauge invariant QED framework by Nambu \cite{nambu} quite a
long time ago\footnote{%
This constraint in the classical electrodynamics framework was originally
suggested by Dirac \cite{dir} (see also \cite{ven} for further developments).%
}, and then in more detail later \cite{az,kep,jej,cj}. The constraint (\ref%
{const}) is in fact very similar to the constraint appearing in the
nonlinear $\sigma $-model for pions \cite{GLA}. It means, in essence, that
the vector field $A_{\mu }$ develops some constant background value and the
Lorentz symmetry $SO(1,3)$ formally breaks down to $SO(3)$ or $SO(1,2)$,
depending on the time-like ($n^{2}>0$) or space-like ($n^{2}<0$) nature of
SLIV. The point is, however, that, in sharp contrast to the nonlinear $%
\sigma $ model for pions, the nonlinear QED theory ensures that all the
physical Lorentz violating effects strictly cancel out among themselves, due
to the starting gauge invariance involved\footnote{%
Let us note, to make things clearer, that the length-fixing vector field
constraint (\ref{const}) is definitely Lorentz invariant by itself.
Nonetheless, as is usual for the nonlinear sigma type models, this
constraint means at the same time the spontaneous Lorentz violation. The
point is, however, that in gauge invariant theories this violation becomes
artificial being converted into gauge degrees of freedom rather than
physical ones. In consequence, ordinary photons and other gauge fields (see
below) appear in essence as the Goldstonic fields that could only be seen
when taking the above nonlinear constraint (nonlinear gauge condition). In
this connection, any other gauge, e.g. Coulomb gauge, is not in line with
Goldstonic picture, since it breaks Lorentz invariance in an explicit rather
than spontaneous way.}.

Furthermore, the most important property of the nonlinear vector field
constraint (\ref{const}) was shown \cite{cj} to be that one does not need to
specially postulate the starting gauge invariance. This was done in the
framework of an arbitrary relativistically invariant Lagrangian containing
adjustable parameters, which is proposed only to possess some global
internal symmetry. Indeed, the SLIV constraint (\ref{const}) causing the
condensation of a generic vector field or vector field multiplet, due to
which the true vacuum in a theory is chosen, happens by itself to be
powerful enough to require adjustment of the parameters to give gauge
invariance. Namely, the existence of the constraint (\ref{const}) is taken
to be upheld by adjusting the parameters of the Lagrangian, in a way that
leads to gauging of the starting global symmetry of the interacting vector
and matter fields involved. In essence, the gauge invariance appears as a
necessary condition for these vector fields not to be superfluously
restricted in degrees of freedom as soon as the SLIV constraint holds.
Indeed, a further reduction in the number of independent $A_{\mu }$
components would make it impossible to set the required initial conditions
in the appropriate Cauchy problem and, in quantum theory, to choose
self-consistent equal-time commutation relations \cite{ogi3}.

Extending the above argumentation, we consider here spontaneous Lorentz
violation realized through a nonlinear length-fixing tensor field constraint
of the type%
\begin{equation}
H_{\mu \nu }H^{\mu \nu }=\mathfrak{n}^{2}M^{2}\text{ , \ \ \ \ }\mathfrak{n}%
^{2}\equiv \mathfrak{n}_{\mu \nu }\mathfrak{n}^{\mu \nu }=\pm 1\text{. }
\label{const3}
\end{equation}
Here $\mathfrak{n}_{\mu \nu }$ is a properly oriented `unit' Lorentz tensor,
while $M$ is the proposed scale for Lorentz violation. We show that such a
type of SLIV induces massless tensor Goldstone modes some of which can
naturally be collected in the physical graviton. The underlying
diffeomophism (diff) invariance appears as a necessary condition for a
symmetric two-tensor field $H_{\mu \nu }$ in Minkowski spacetime not to be
superfluously restricted in degrees of freedom, apart from the constraint
due to which the true vacuum in a theory is chosen by the Lorentz violation.

\subsection{Outline of the paper}

The paper is organized in the following way. Further in this section we
discuss the main features of SLIV regarding both input and emergent gauge
invariance. The focus of this paper will be on emergent gauge invariance. In
section 2 we review the emergent QED and Yang-Mills theories \cite{cj},
which appear due to a SLIV constraint being put on a vector field or a
vector field multiplet, respectively. In section 3 we generalize this
approach to the tensor field case and find the emergent gravity theory whose
vacuum is also determined by spontaneous Lorentz violation. Finally, in
section 4, we present a r\'{e}sum\'{e} and conclude.

\subsection{SLIV: an intact physical Lorentz invariance}

The original models realizing the SLIV conjecture were based on a four
fermion (current-current) interaction, where the massless vector NG modes
appear as fermion-antifermion pair composite states \cite{bjorken}. This is
in complete analogy with the massless composite scalar modes in the original
Nambu-Jona-Lazinio model \cite{NJL}. Unfortunately, owing to the lack of a
starting gauge invariance in such models and the composite nature of the NG
modes which appear, it is hard to explicitly demonstrate that these modes
together really form a massless vector boson as a gauge field candidate
universally interacting with all kinds of matter. Rather, there are in
general three separate massless NG modes, two of which may mimic the
transverse photon polarizations, while the third one must be appropriately
suppressed.

In this connection, the more instructive laboratory for SLIV consideration
proves to be a simple class of QED type models \cite{nambu} having from the
outset a gauge invariant form, in which the spontaneous Lorentz violation is
realized through the nonlinear constraint (\ref{const}). Remarkably, this
type of model makes the vector Goldstone boson a true gauge boson (photon),
whereas the physical Lorentz invariance is left intact. Indeed, despite an
evident similarity with the nonlinear $\sigma $-model for pions, the
nonlinear QED theory ensures that all the physical Lorentz violating effects
prove to be non-observable, due to the starting gauge invariance involved.
It was shown \cite{nambu}, while only in the tree approximation and for
time-like SLIV ($n^{2}>0$), that the non-linear constraint (\ref{const})
implemented as a supplementary condition into the standard QED Lagrangian
containing the charged fermion field $\psi (x)$ 
\begin{equation}
L_{QED}=-\frac{1}{4}F_{\mu \nu }F^{\mu \nu }+\overline{\psi }(i\gamma
\partial +m)\psi -eA_{\mu }\overline{\psi }\gamma ^{\mu }\psi \text{ , \ }%
A_{\mu }A^{\mu }=n^{2}M^{2}\text{\ }  \label{lag11}
\end{equation}%
appears in fact as a possible gauge choice for the vector field $A_{\mu }$.
At the same time the $S$-matrix remains unaltered under such a gauge
convention. Really, this nonlinear QED contains a plethora of Lorentz and $%
CPT$ violating couplings when it is expressed in terms of the pure
Goldstonic photon modes ($a_{\mu }$) according to the constraint condition (%
\ref{const}) 
\begin{equation}
A_{\mu }=a_{\mu }+\frac{n_{\mu }}{n^{2}}(M^{2}-n^{2}a^{2})^{\frac{1}{2}}%
\text{ , \ }n_{\mu }a_{\mu }=0\text{ \ \ \ \ (}a^{2}\equiv a_{\mu }a^{\mu }%
\text{).}  \label{gol}
\end{equation}%
In addition there is an effective \textquotedblleft Higgs" mode $(n_{\mu
}/n^{2})(M^{2}-n^{2}a^{2})^{1/2}$ given by the constraint (for definiteness,
one takes the positive sign for the square root when expanding it in powers
of $a^{2}/M^{2}$). However, the contributions of these Lorentz violating
couplings to physical processes completely cancel out among themselves. So,
SLIV was shown to be superficial as it affects only the gauge of the vector
potential $A_{\mu }$, at least in the tree approximation \cite{nambu}.

Some time ago, this result was extended to the one-loop approximation and
for both time-like ($n^{2}>0$) and space-like ($n^{2}<0$) Lorentz violation 
\cite{az}. All the contributions to the photon-photon, photon-fermion and
fermion-fermion interactions violating physical Lorentz invariance were
shown to exactly cancel among themselves, in the manner observed by Nambu
long ago for the simplest tree-order diagrams. This means that the
constraint (\ref{const}), having been treated as a nonlinear gauge choice at
the tree (classical) level, remains as a gauge condition when quantum
effects are taken into account as well. So, in accordance with Nambu's
original conjecture, one can conclude that physical Lorentz invariance is
left intact at least in the one-loop approximation, provided that we
consider the standard gauge invariant QED Lagrangian (\ref{lag11}) taken in
flat Minkowski spacetime. Later this result was also confirmed for
spontaneously broken massive QED \cite{kep}, non-Abelian theories \cite{jej}
and tensor field gravity \cite{cjt}. Some interesting aspects of SLIV in
nonlinear QED were considered in \cite{ur}.

\subsection{SLIV: emergent gauge symmetries}

In the above-discussed models, due to the assumed gauge symmetry, physical
Lorentz invariance always appears intact, in the sense that all Lorentz
non-invariant effects caused by the vector field vacuum expectation values
(vevs) are physically unobservable. However the most important property of
the nonlinear vector field SLIV constraint (\ref{const}), was shown \cite{cj}
to be that one does not have to impose gauge symmetry directly. Indeed we
showed that gauge invariance was unavoidable, if the equations of motion
should have enough freedom to allow a constraint like (\ref{const}) to be
fulfilled. This need for gauge symmetry was deduced in a model with the
nonlinear $\sigma $-model type spontaneous Lorentz violation, in the
framework of an arbitrary relativistically invariant Lagrangian for
elementary vector and matter fields, which are proposed only to possess some
global internal symmetry. One simply assumes that the existence of the
constraint (\ref{const}) is to be upheld by adjusting the parameters of the
Lagrangian. The SLIV conjecture happens to be powerful enough by itself to
require gauge invariance, provided that we allow the parameters in the
corresponding Lagrangian density to be adjusted so as to ensure
self-consistency without losing too many degrees of freedom. Namely, due to
the spontaneous Lorentz violation determined by the constraint (\ref{const}%
), the true vacuum in such a theory is chosen so that this theory acquires
on its own a gauge-type invariance, which gauges the starting global
symmetry of the interacting vector and matter fields involved. In essence,
the gauge invariance (with a proper gauge-fixing term) appears as a
necessary condition for these vector fields not to be superfluously
restricted in degrees of freedom.

Let us dwell upon this point in more detail. Generally, while a conventional
variation principle requires the equations of motion to be satisfied, it is
possible to eliminate one component of a general 4-vector field $A_{\mu }$,
in order to describe a pure spin-1 particle by imposing a supplementary
condition. In the massive vector field case there are three physical spin-1
states to be described by the $A_{\mu }$ field. Similarly in the massless
vector field case, although there are only two physical (transverse) photon
spin states, one cannot construct a massless 4-vector field $A_{\mu }$ as a
linear combination of creation and annihilation operators for helicity $\pm
1 $ states in a relativistically covariant way, unless one fictitious state
is added \cite{GLB}. So, in both the massive and massless vector field
cases, only one component of the $A_{\mu }$ field may be eliminated and
still preserve Lorentz invariance. Once the SLIV constraint (\ref{const}) is
imposed, it is therefore not possible to satisfy another supplementary
condition, since this would superfluously restrict the number of degrees of
freedom for the vector field. In fact a further reduction in the number of
independent $A_{\mu }$ components would make it impossible to set the
required initial conditions in the appropriate Cauchy problem and, in
quantum theory, to choose self-consistent equal-time commutation relations 
\cite{ogi3}.

We now turn to the question of the consistency of a constraint with the
equations of motion for a general 4-vector field $A_{\mu }$ Actually, there
are only two possible covariant constraints for such a vector field in a
relativistically invariant theory - the holonomic SLIV constraint, $%
C(A)=A_{\mu }A^{\mu }-n^{2}M^{2}=0$ (\ref{const}), and the non-holonomic
one, known as the Lorentz condition, $C(A)=\partial _{\mu }A^{\mu }=0$. In
the presence of the SLIV constraint $C(A)=A^{\mu }A_{\mu }-n^{2}M^{2}=0$, it
follows that the equations of motion can no longer be independent. The
important point is that, in general, the time development would not preserve
the constraint. So the parameters in the Lagrangian have to be chosen in
such a way that effectively we have one less equation of motion for the
vector field. This means that there should be some relationship between all
the (vector and matter) field Eulerians ($E_{A}$, $E_{\psi }$, ...) involved%
\footnote{$E_{A}$ stands for the vector-field Eulerian $(E_{A})^{\mu }\equiv
\partial L/\partial A_{\mu }-\partial _{\nu }[\partial L/\partial (\partial
_{\nu }A_{\mu })].$ We use similar notations for other field Eulerians as
well.}. Such a relationship can quite generally be formulated as a
functional - but by locality just a function - of the Eulerians, $%
F(E_{A},E_{\psi },...)$, being put equal to zero at each spacetime point
with the configuration space restricted by the constraint $C(A)=0$: 
\begin{equation}
F(C=0;\text{ \ }E_{A},E_{\psi },...)=0\text{ .}  \label{FF}
\end{equation}%
This relationship must satisfy the same symmetry requirements of Lorentz and
translational invariance, as well as all the global internal symmetry
requirements, as the general starting Lagrangian $L(A,\psi ,...)$ does. We
shall use this relationship in subsequent sections as the basis for gauge
symmetry generation in the SLIV constrained vector and tensor field theories.

Let us now consider a \textquotedblleft Taylor expansion" of the function F
expressed as a linear combination of terms involving various field
combinations multiplying or derivatives acting on the Eulerians\footnote{%
The Eulerians are of course just particular field combinations themselves
and so this \textquotedblleft expansion" at first includes higher powers and
higher derivatives of the Eulerians.}. The constant term in this expansion
is of course zero since the relation (\ref{FF}) must be trivially satisfied
when all the Eulerians vanish, i.e. when the equations of motion are
satisfied. We now consider just the terms containing field combinations (and
derivatives) with mass dimension 4, corresponding to the Lorentz invariant
expressions 
\begin{equation}
\partial _{\mu }(E_{A})^{\mu },\text{ }A_{\mu }(E_{A})^{\mu },\text{ }%
E_{\psi }\psi ,\text{ }\overline{\psi }E_{\overline{\psi }}.  \label{fff}
\end{equation}%
All the other terms in the expansion contain field combinations and
derivatives with higher mass dimension and must therefore have coefficients
with an inverse mass dimension. We expect the mass scale associated with
these coefficients should correspond to a large fundamental mass (e.g. the
Planck mass $M_{P}$). Hence we conclude that such higher dimensional terms
must be highly suppressed and can be neglected. A priori these neglected
terms could lead to the breaking of the spontaneously generated gauge
symmetry at high energy. However it could well be that a more detailed
analysis would reveal that the imposed SLIV constraint requires an exact
gauge symmetry. Indeed, if one uses classical equations of motion, a gauge
breaking term will typically predict the development of the
\textquotedblleft gauge\textquotedblright\ in a way that is inconsistent
with our gauge fixing constraint $C(A)=0$. Thus the theory will generically
only be consistent if it has exact gauge symmetry.

In the above discussion we have simply considered a single vector field.
However in sections 2 and 3 we shall also consider a non-Abelian vector
field $A_{\mu }^{a}$ and a tensor field $H_{\mu \nu }$ respectively. In
these cases the lowest mass dimension terms analogous to the expressions (%
\ref{fff}) have symmetry indices. The function analogous to $F$ in equation (%
\ref{FF}), which is a linear combination of these terms, must respect the
assumed global non-Abelian symmetry and Lorentz symmetry. So all the terms
must transform in the same way and carry the same symmetry index, $a$ or $%
\nu $ respectively, which is then inherited by the function analogous to $F$%
. Since gravitational interactions vanish in the low energy limit, we have
to include dimension 5 terms in our function $\mathcal{F}^{\mu }$ for the
gravity case.

The other possible Lorentz covariant constraint $\partial _{\mu }A^{\mu }=0,$
while also being sensitive to the form of the constraint-compatible
Lagrangian, leads to massive QED and massive Yang-Mills theories \cite{ogi3}.

In the case of a symmetric two-tensor field $H_{\mu \nu }$, we consider
spontaneous Lorentz violation realized through a nonlinear tensor field
constraint of the type (\ref{const3}). This constraint fixes the length of
the tensor field in an analogous way to that of the vector field above by
the constraint (\ref{const}). For consistency between this constraint (\ref%
{const3}) and the equations of motion, we require the parameters of the
theory to be chosen in such a way that the above-mentioned relationship $%
\mathcal{F}^{\mu }=0$ be satisfied. As a result, the theory turns out to
correspond to linearized general relativity in the weak field approximation,
while some of the massless tensor Goldstone modes appearing through SLIV are
naturally collected in the physical graviton. The accompanying diffeomophism
invariance appears as a necessary condition for the symmetric two-tensor
field $H_{\mu \nu }$ in Minkowski spacetime not to be superfluously
restricted in degrees of freedom, apart from the constraint due to which the
true vacuum in the theory is chosen by the Lorentz violation. The emergent
theory looks essentially nonlinear when expressed in terms of the pure
Goldstone tensor modes and contains, besides general relativity (GR) in the
weak-field limit approximation, a variety of new Lorentz and $CPT$ violating
couplings. However, they do not lead to physical Lorentz violation, due to
the simultaneously generated diffeomophism invariance, once the tensor field
gravity theory (being considered as the weak-field limit of general
relativity) is properly extended to GR\footnote{%
Remarkably, the diff invariance appears so powerful that not only
spontaneous but even explicit Lorentz violation may sometimes be converted
into gauge degrees of freedom. One interesting example \cite{jiv} is related
to Chern-Simons modified gravity where the apparent Lorentz symmetry
breaking may in fact be just a choice of gauge.}. So, this formulation of
SLIV seems to amount to the fixing of a gauge for the tensor field in a
special manner, making the Lorentz violation only superficial just as in the
nonlinear QED framework \cite{nambu}. From this viewpoint, both conventional
QED and GR theories appear to be Goldstonic theories, in which some of the
gauge degrees of freedom of these fields are condensed and eventually emerge
as a noncovariant gauge choice. The associated massless NG modes are
collected in photons and gravitons, in such a way that physical Lorentz
invariance is ultimately preserved.

\section{The Vector Goldstone Boson Primer}

\subsection{Emergent QED}

Let us consider an arbitrary relativistically invariant Lagrangian $L(A,\psi
)$ of one vector field $A_{\mu }$ and one complex matter field $\psi $,
taken to be a charged fermion for definiteness, in an Abelian model with the
corresponding global $U(1)$ charge symmetry imposed. For convenience and the
apparent simplicity of the method, we choose to impose the SLIV constraint (%
\ref{const}) using a well-known classical procedure for holonomic
constraints (see, for example, \cite{lan}), involving a Lagrange multiplier
term in an appropriately extended Lagrangian $L^{\prime}(A,\psi, \lambda )$.
Since the main point of the present article is to consider theories that
become inconsistent unless they have special relations between the
parameters of the theory -- making them into gauge theories -- we want to
impose the SLIV constraint in a way that leads generically to such an
inconsistency. The trick we use to achieve this is to arrange for the
Lagrange multiplier field $\lambda (x)$ to disappear from the equations of
motion (Eulerians) for the other fields. In order that the auxiliary field $%
\lambda (x)$, which acts as the Lagrange multiplier, should not appear in
the equations of motion, we take a quadratic form for the Lagrange
multiplier term as follows 
\begin{equation}
L^{\prime }(A,\psi ,\lambda )=L(A,\psi )-\frac{1}{4}\lambda \left( A_{\nu
}A^{\nu }-n^{2}M^{2}\right) ^{2}.  \label{qu}
\end{equation}%
Varying $L^{\prime }(A,\psi, \lambda )$ with respect to the auxiliary field $%
\lambda (x)$ gives the equation of motion 
\begin{equation}
E_{\lambda }^{\prime }=\partial L^{\prime }/\partial \lambda =\frac{1}{4}%
\left( A_{\nu }A^{\nu }-n^{2}M^{2}\right) ^{2}=0,  \label{on11}
\end{equation}%
leading to just the SLIV condition (\ref{const}). The equations of motion
for $A_{\mu }$ in this case are independent of the $\lambda (x)$, which
completely decouples from them rather than acting as some extra source of
charge density, as it would in the case of a linear Lagrange multiplier term%
\footnote{%
Indeed, in this case one could propose that the auxiliary field $\lambda (x)$
is chosen in such a way that this extra source current is conserved $%
\partial _{\mu }(\lambda A^{\mu })=0$, according to which if the auxiliary
field $\lambda (x)$\ is fixed at one instant of time its value at other
times can be then determined by this conservation law. Otherwise, with an
arbitrary $\lambda (x),$ this field could have an uncontrollable influence
on the vector field dynamics. However, this conservation law would in fact
constitute an additional condition on the theory since, in contrast to a
conventional Noether fermion current in the starting $U(1)$ globally
invariant Lagrangian $L(A,\psi )$, this current $j^{\mu }=\lambda A^{\mu }$
is not automatically conserved.}.

Now, under the assumption that the SLIV constraint is preserved under the
time development given by the equations of motion, we show how gauge
invariance of the starting Lagrangian $L(A,\psi )$ is established. A
conventional variation principle applied to the total Lagrangian $L^{\prime
}(A,\psi ,\lambda )$ requires the following equations of motion for the
vector field $A_{\mu }$ and the auxiliary field $\lambda $ to be satisfied%
\begin{equation}
(E_{A}^{\prime })^{\mu }=(E_{A})^{\mu }=0\text{ , \ \ \ }C(A)=A_{\nu }A^{\nu
}-n^{2}M^{2}=0\text{\ },  \label{em}
\end{equation}%
where the Eulerian $(E_{A})^{\mu }$ is given by the starting Lagrangian $%
L(A,\psi )$. However, in accordance with the general argumentation given in
the Introduction, the existence of five equations for the 4-component vector
field $A^{\mu }$ (one of which is the constraint) means that not all of the
vector field Eulerian components can be independent. Therefore, there must
be a relationship of the form $F(C=0;\text{ \ }E_{A},E_{\psi },...)=0$ given
in equation (\ref{FF}), expressed as a linear combination of the dimension 4
Lorentz invariant expressions given in equation (\ref{fff}). It follows that
the parameters in the Lagrangian $L(A,\psi )$ must be chosen so as to
satisfy an identity between the vector and matter field Eulerians of the
following type 
\begin{equation}
\partial _{\mu }(E_{A})^{\mu }=cA_{\mu }(E_{A})^{\mu }+itE_{\psi }\psi -it%
\overline{\psi }E_{\overline{\psi }}.  \label{div}
\end{equation}%
This identity immediately signals the invariance of the basic Lagrangian $%
L(A,\psi )$ under vector and fermion field local transformations whose
infinitesimal form is given by\footnote{%
Since the Eulerians are functional derivatives of the action, e.g.~$%
(E_{A})^{\mu }=\frac{\delta S}{\delta A_{\mu }}$, a relation such as (\ref%
{div}) between them implies that a certain combined variation of the various
fields with the variations $\delta A_{\mu }$, $\delta \psi $,.. being
proportional to the corresponding coefficients $cA_{\mu }$, $it\psi $,.. of
the Eulerians in (\ref{div}) does not change $S$.} 
\begin{equation}
\delta A_{\mu }=\partial _{\mu }\omega +c\omega A_{\mu },\text{ \ \ }\delta
\psi \text{\ }=it\omega \psi \text{ .}  \label{trans}
\end{equation}%
Here $\omega (x)$ is an arbitrary function, only being restricted by the
requirement to conform with the nonlinear constraint (\ref{const})%
\begin{equation}
(A_{\mu }+\partial _{\mu }\omega +c\omega A_{\mu })(A^{\mu }+\partial ^{\mu
}\omega +c\omega A_{\mu })=n^{2}M^{2}\text{ .}
\end{equation}%
Conversely, the identity (\ref{div}) follows from the invariance of the
Lagrangian $L(A_{\mu },\psi )$ under the transformations (\ref{trans}).
Indeed, both direct and converse assertions are particular cases\footnote{%
In general Noether's theorem applies to the invariance of an action rather
than the invariance of a Lagrangian. However these are both completely
equivalent, unless one considers spacetime symmetries with a local variation
of coordinates as well (see section 3).} of Noether's second theorem \cite%
{noeth}. The point is, however, that these transformations cannot in general
form a group unless the constant $c$ vanishes. In fact, by constructing the
corresponding Lie bracket operation $(\delta _{1}\delta _{2}-\delta
_{2}\delta _{1})$ for two successive vector field variations we find that,
while the fermion transformation in (\ref{trans}) is an ordinary Abelian
local one with zero Lie bracket, for the vector field transformations there
appears a non-zero result 
\begin{equation}
(\delta _{1}\delta _{2}-\delta _{2}\delta _{1})A_{\mu }=c(\omega
_{1}\partial _{\mu }\omega _{2}-\omega _{2}\partial _{\mu }\omega _{1}),
\label{SL1}
\end{equation}%
which is proportional to the constant $c$. Thus we necessarily require $c=0$
for the bracket operation to be closed. Note also that for non-zero $c$ the
variation of $A_{\mu }$ given by (\ref{SL1}) is an essentially arbitrary
vector function. Such a freely varying $A_{\mu }$ is only consistent with a
trivial Lagrangian (i.e. $L=const$). Thus, in order to have a non-trivial
Lagrangian, it is necessary to have $c=0$ and the theory given by the basic
Lagrangian $L(A_{\mu },\psi )$ then possesses an Abelian local symmetry%
\footnote{%
We shall see below that non-zero $c$-type coefficients appear in the
non-Abelian internal symmetry case, resulting eventually in a Yang-Mills
gauge invariant theory.}. \ 

We have now shown how the choice of a vacuum conditioned by the SLIV
constraint (\ref{const}) enforces the choice of the parameters in the
starting Lagrangian $L(A_{\mu },\psi )$, so as to convert the starting
global $U(1)$ charge symmetry into a local one. This SLIV induced local
Abelian symmetry (\ref{trans}) allows the total Lagrangian $L^{\prime }$ to
be determined in full. For a theory with renormalizable coupling constants,
it is in fact the conventional QED Lagrangian (\ref{lag11}) extended by the
Lagrange multiplier term, which provides the SLIV constraint (\ref{const})
imposed on the vector field $A_{\mu }.$ Thus, we eventually come to the
total Lagrangian%
\begin{equation}
L^{\prime }(A,\psi ,\lambda )=L_{QED}-\frac{1}{4}\lambda \left( A_{\mu
}A^{\mu }-n^{2}M^{2}\right) ^{2}  \label{qed2}
\end{equation}%
in the most direct way. This type of Abelian vector field theory with a
quadratic Lagrange multiplier term was recently considered in \cite{kkk}.
The equations of motion generated by this theory are the equations in the
absence of the constraint (\ref{const}) plus the constraint itself. Thus the
introduction of the quadratic Lagrange-multiplier type of term is in fact
equivalent at the classical level to imposing the constraint on the
equations of motion by hand\footnote{%
Just the latter approach was used in our previous analysis \cite{cj} of
gauge symmetry generation in SLIV constrained vector field theories. Here we
follow the variational treatment of this constraint, although the only
distinction between the two approaches is the presence of a decoupled
Lagrange-multiplier field $\lambda (x)$ which is actually left undetermined
in the theory.}. This theory is closely related to the Nambu QED model (\ref%
{lag11}), in which the SLIV constraint is proposed to be substituted into
the Lagrangian before varying the action, although the correspondence is not
exact. The Nambu model yields a total of four equations for the fields: the
constraint by itself and three equations of motion from the variation.
Meanwhile, the model (\ref{qed2}) yields five equations of motion instead,
one of which is the constraint. The extra equation corresponds to the Gauss
law, which in the Nambu approach is imposed as a separate initial condition
that subsequently holds at all times, by virtue of the three equations of
motion and the constraint \cite{kkk}. They both lead to SLIV, which
generates massless Goldstone modes associated with photons and forces the
massive mode to vanish. This pattern of SLIV emerges as a noncovariant gauge
choice in an otherwise gauge invariant and Lorentz invariant theory, as was
already discussed in the Introduction.

\subsection{Emergent Yang-Mills theories}

We shall here discuss the non-Abelian internal symmetry case and show that
the Yang-Mills gauge fields also appear as possible vector Goldstone modes,
when the true vacuum in the theory is chosen by the non-Abelian SLIV
constraint 
\begin{equation}
Tr(\boldsymbol{A}_{\mu }\boldsymbol{A}^{\mu })=\boldsymbol{n}^{2}M^{2},\text{
\ \ }\boldsymbol{n}^{2}\equiv \boldsymbol{n}_{\mu }^{a}\boldsymbol{n}^{\mu
,a}=\pm 1,  \label{const2}
\end{equation}%
where $\boldsymbol{n}_{\mu }^{a}$ is now some `unit' rectangular matrix. We
consider a general Lorentz invariant Lagrangian $\mathrm{L}(\boldsymbol{A}%
_{\mu },\boldsymbol{\psi })$ for the vector and matter fields involved
possessing some global internal symmetry given by a group $G$ with $D$
generators $t^{a}$ 
\begin{equation}
\lbrack t_{a},t_{b}]=ic_{abc}t_{c},\text{ \ }Tr(t_{a}t_{b})=\delta _{ab}%
\text{ \ \ }(a,b,c=0,1,...,D-1),  \label{com}
\end{equation}%
where $c_{abc}$ are the structure constants of $G$. The corresponding vector
fields, which transform according to the adjoint representation of $G$, are
given in the matrix form $\boldsymbol{A}_{\mu }=\boldsymbol{A}_{\mu
}^{a}t_{a}$. The matter fields (fermion fields for definiteness) are taken
in the fundamental representation column $\boldsymbol{\psi }^{\sigma }$ ($%
\sigma =0,1,...,d-1$) of $G$.

We impose the SLIV constraint (\ref{const2}), as in the above Abelian case,
by introducing an extended Lagrangian $\mathrm{L}^{\prime }$ containing a
quadratic Lagrange multiplier term 
\begin{equation}
\mathrm{L}^{\prime }(\boldsymbol{A}_{\mu },\boldsymbol{\psi }, \lambda )=%
\mathrm{L}(\boldsymbol{A}_{\mu },\boldsymbol{\psi })- \lambda/4[Tr(%
\boldsymbol{A}_{\mu }\boldsymbol{A}^{\mu })-\boldsymbol{n}^{2}M^{2}]^{2}.
\label{tot}
\end{equation}%
The variation of $\mathrm{L}^{\prime}(\boldsymbol{A}_{\mu },\boldsymbol{\psi 
}, \lambda)$ with respect to $\boldsymbol{A}_{\mu }$ gives the vector field
equation of motion 
\begin{equation}
(\mathrm{E}_{\boldsymbol{A}})_{a}^{\mu }-\lambda\boldsymbol{A}_{a}^{\mu }[Tr(%
\boldsymbol{A}_{\mu }\boldsymbol{A}^{\mu })-\boldsymbol{n}^{2}M^{2}]=0\text{
\ \ \ }(a=0,1,...,D-1).  \label{eqmI}
\end{equation}%
Here the vector field Eulerian $\mathrm{E}_{\boldsymbol{A}}$ is determined
by the starting Lagrangian $\mathrm{L}(\boldsymbol{A}_{\mu },\boldsymbol{%
\psi })$, while the Eulerian of the auxiliary field $\lambda(x) $ taken
on-shell%
\begin{equation}
\mathrm{E}_{\lambda }^{\prime }=\partial \mathrm{L}^{\prime }/\partial
\lambda =\frac{1}{4}[Tr(\boldsymbol{A}_{\mu }\boldsymbol{A}^{\mu })-%
\boldsymbol{n}^{2}M^{2}]^{2}=0\text{ \ \ \ }  \label{on2}
\end{equation}%
gives the constraint (\ref{const2}). So, once the constraint holds, one has
the following simplified equations for the vector fields 
\begin{equation}
(\mathrm{E}_{\boldsymbol{A}})_{a}^{\mu }=0\text{ , \ \ } C(\boldsymbol{A}%
_{\mu }) = Tr(\boldsymbol{A}_{\mu }\boldsymbol{A}^{\mu }) -\boldsymbol{n}%
^{2}M^{2}=0,  \label{eqs}
\end{equation}%
whereas the auxiliary field $\lambda(x),$ as in the Abelian case, entirely
decouples from the vector field dynamics.

The need to preserve the constraint $C(\boldsymbol{A}_{\mu }) = 0$ with time
implies that the equations of motion for the vector fields $\boldsymbol{A}%
_{\mu }^{a}$ cannot be all independent. Consequently the parameters in the
Lagrangian $\mathrm{L}(\boldsymbol{A}_{\mu },\boldsymbol{\psi })$ must be
chosen so as to give a relationship between the Eulerians for the vector and
matter fields analogous to equation (\ref{FF}). We include just the lowest
dimensional Lorentz invariant expressions constructed from the Eulerians in
this relationship, on the grounds that other terms will be suppressed by a
large mass parameter like $M_P$. These lowest dimension terms include $%
\partial _{\mu }(\mathrm{E}_{\boldsymbol{A}})_{a}^{\mu }$ and all the terms
in the relationship must transform in the same way under the global symmetry
group G. Hence the relationship must transform as the adjoint representation
of G and carry the symmetry index $a$ 
\begin{equation}
\boldsymbol{F}_{a}(C=0;\text{ \ }\mathrm{E}_{\boldsymbol{A}},\mathrm{E}_{%
\boldsymbol{\psi }},...)=0\text{ \ \ \ }(a=0,1,...,D-1).  \label{FF1}
\end{equation}
It therefore takes the following form 
\begin{equation}
\partial _{\mu }(\mathrm{E}_{\boldsymbol{A}})_{a}^{\mu }= d_{\boldsymbol{A}%
}c_{abc}\boldsymbol{A}_{\mu }^{b}(\mathrm{E}_{\boldsymbol{A}})^{\mu ,c} + d_{%
\boldsymbol{\psi}}\mathrm{E}_{\boldsymbol{\psi }} (it_{a})\boldsymbol{\psi }%
+ d_{\overline{\boldsymbol{\psi }}}\overline{\boldsymbol{\psi }}(-it_{a})%
\mathrm{E}_{\overline{\boldsymbol{\psi }}},  \label{id11}
\end{equation}
where $d_{\boldsymbol{A}}$, $d_{\boldsymbol{\psi}}$ and $d_{\overline{%
\boldsymbol{\psi }}}$ are as yet undetermined constants. Noether's second
theorem \cite{noeth} can be applied directly to this identity (\ref{id11}),
in order to derive the invariance of $\mathrm{L}(\boldsymbol{A}_{\mu },%
\boldsymbol{\psi })$ under vector and fermion field local transformations
having the infinitesimal form 
\begin{equation}
\delta \boldsymbol{A}_{\mu }^{a}=\partial _{\mu }\omega ^{a}+ d_{\boldsymbol{%
A}}c_{abc}\boldsymbol{A}_{\mu }^{b}\omega^{c}, \text{ \ \ }\delta 
\boldsymbol{\psi } \text{\ }= d_{\boldsymbol{\psi}}(it_{a})\omega ^{a}%
\boldsymbol{\psi } ,\text{ \ \ }\delta \overline{\boldsymbol{\psi }}\text{\ }%
= d_{\overline{\boldsymbol{\psi }}}\overline{\boldsymbol{\psi }}
(-it_{a})\omega ^{a}.  \label{trans1}
\end{equation}

Of course from the symmetry transformations (\ref{trans1}) one can generate
the commutators $(\delta_1\delta_2-\delta_2\delta_1)\boldsymbol{A}_{\mu
}^{a} $, $(\delta_1\delta_2-\delta_2\delta_1)\boldsymbol{\psi}$ and $%
(\delta_1\delta_2-\delta_2\delta_1)\overline{\boldsymbol{\psi }}$ as new
symmetry transformations. However, in order to avoid generating too many
symmetry transformations which would essentially only be consistent with the
Lagrangian density being a constant, we need that the Lie algebra of the
transformations should close. That is to say we need relations between the
above Lie brackets of the form 
\begin{equation}
(\delta_1\delta_2-\delta_2\delta_1) = \delta_{br},  \label{lb}
\end{equation}
where the functions $\omega_{br}^a(x)$ associated with the transformation $%
\delta_{br}$ are expressed in terms of the functions $\omega_1^a(x)$ and $%
\omega_2^a(x)$ for the transformations $\delta_1$ and $\delta_2$. For
example 
\begin{equation}
(\delta_1\delta_2-\delta_2\delta_1)\boldsymbol{\psi} = d_{\boldsymbol{\psi}%
}^2[it_a,it_b]\omega_1^a\omega_2^b\boldsymbol{\psi}  \label{lbpsi}
\end{equation}
can be interpreted as 
\begin{equation}
\delta_{br}\boldsymbol{\psi} = d_{\boldsymbol{\psi}}it_c\omega_{br}^c 
\boldsymbol{\psi}  \label{brpsi}
\end{equation}
provided that 
\begin{equation}
\omega_{br}^c=-d_{\boldsymbol{\psi}}c_{abc}\omega_1^a\omega_2^b.
\label{ombrpsi}
\end{equation}
Corresponding formulas apply for the Lie bracket of two symmetry
transformations acting on $\overline{\boldsymbol{\psi }}$ with 
\begin{equation}
\omega_{br}^c=-d_{\overline{\boldsymbol{\psi }}} c_{abc}\omega_1^a\omega_2^b.
\label{ombrpsibar}
\end{equation}
Similarly the Lie bracket for the $\boldsymbol{A}_{\mu}^a$ field is given by 
\begin{equation}
(\delta_1\delta_2-\delta_2\delta_1)\boldsymbol{A}_{\mu}^a = -d_{\boldsymbol{A%
}}c_{abc}\partial_{\mu}(\omega_1^b\omega_2^c) + d_{\boldsymbol{A}%
}^2c_{abc}c_{bde}(\omega_1^c\omega_2^e- \omega_2^c\omega_1^e)\boldsymbol{A}%
_{\mu}^d.
\end{equation}
Using the Jacobi identity, we then obtain the closure of the Lie algebra on
the $\boldsymbol{A}_{\mu}^a$ field with 
\begin{equation}
\omega_{br}^c=-d_{\boldsymbol{A}}c_{abc}\omega_1^a\omega_2^b.  \label{ombra}
\end{equation}
In order to obtain full closure of the Lie algebra for all the fields, we
require that the three expressions (\ref{ombrpsi}), (\ref{ombrpsibar}) and (%
\ref{ombra}) for $\omega_{br}^c$ should be identical. Thus we obtain 
\begin{equation}
d_{\boldsymbol{A}} = d_{\boldsymbol{\psi}} = d_{\overline{\boldsymbol{\psi}}%
}.  \label{deq}
\end{equation}
Here the $\omega ^{a}(x)$ are arbitrary functions only being restricted,
again as in the above Abelian case, by the requirement to conform with the
corresponding nonlinear constraint (\ref{const2}).

So, by choosing the parameters in the Lagrangian to be consistent with the
constraint (\ref{const2}), we have obtained a non-Abelian gauge symmetry
under the transformations (\ref{trans1}) with the coefficients satisfying (%
\ref{deq}). In order to construct a non-Abelian field tensor $\boldsymbol{F}%
_{\mu\nu}^a$ having the usual relationship 
\begin{equation}
\boldsymbol{F}_{\mu\nu}^a=\partial_{\mu}\boldsymbol{A}_{\nu}^a
-\partial_{\nu}\boldsymbol{A}_{\mu}^a +c_{abc}\boldsymbol{A}_{\mu}^b%
\boldsymbol{A}_{\nu}^c  \label{fmunu}
\end{equation}
with the gauge fields, we have to rescale $\boldsymbol{A}_{\mu}^a$ and $%
\omega^a$ by a factor of $d_{\boldsymbol{A}}^{-1}$ 
\begin{equation}
\boldsymbol{A}_{\mu}^a \rightarrow \frac{\boldsymbol{A}_{\mu}^a}{d_{%
\boldsymbol{A}}}, \text{ \ \ } \omega^a \rightarrow \frac{\omega^a}{d_{%
\boldsymbol{A}}}.  \label{rescale}
\end{equation}
Then the transformations (\ref{trans1}) expressed in terms of the rescaled
field (\ref{rescale}) become the standard non-Abelian gauge transformations.
For a theory with renormalizable coupling constants, this derived gauge
symmetry leads to the conventional Yang-Mills type Lagrangian 
\begin{equation}
\mathrm{L}(\boldsymbol{A}_{\mu },\psi )=-\frac{1}{4g^2}\,Tr(\boldsymbol{F}%
_{\mu \nu }\boldsymbol{F}^{\mu \nu })+\overline{\boldsymbol{\psi }}(i\gamma
\partial -m)\boldsymbol{\psi }+\overline{\boldsymbol{\psi }}\boldsymbol{A}%
_{\mu }\gamma ^{\mu }\boldsymbol{\psi }  \label{nab1}
\end{equation}%
with an arbitrary gauge coupling constant $g$.

Let us turn now to the spontaneous Lorentz violation which is caused by the
nonlinear vector field constraint (\ref{const2}). Although the Lagrangian $%
\mathrm{L}(\boldsymbol{A}_{\mu },\boldsymbol{\psi })$ only has an $%
SO(1,3)\times G$ invariance, the chosen SLIV constraint (\ref{const2})
possesses a much higher accidental symmetry $SO(D,3D)$ determined by the
dimensionality $D$ of the adjoint representation of $G$ to which the vector
fields $\boldsymbol{A}_{\mu }^{a}$ belong. This symmetry is spontaneously
broken at a scale $M,$ together with the actual $SO(1,3)\otimes G$ symmetry,
by the vev 
\begin{equation}
<\boldsymbol{A}_{\mu }^{a}(x)>\text{ }=\boldsymbol{n}_{\mu }^{a}M.
\label{vevv}
\end{equation}%
Here the vacuum direction is now given by the matrix $\boldsymbol{n}_{\mu
}^{a} $ describing simultaneously both of the generalized SLIV cases,
time-like ($SO(D,3D)$ $\rightarrow SO(D-1,3D)$) or space-like ($SO(D,3D)$ $%
\rightarrow SO(D,3D-1)$) respectively, depending on the sign of $\boldsymbol{%
n}^{2}\equiv \boldsymbol{n}_{\mu }^{a}\boldsymbol{n}^{\mu ,a}=\pm 1$. In
both cases this matrix has only one non-zero element, subject to the
appropriate $SO(1,3)$ and (independently) $G$ rotations. They are,
specifically, $\boldsymbol{n}_{0}^{0}$ or $\boldsymbol{n}_{3}^{0}$ provided
that the vacuum expectation value (\ref{vevv}) is developed along the $a=0$
direction in the internal space and along the $\mu =0$ or $\mu =3$ direction
respectively in the ordinary four-dimensional spacetime. Side by side with
one true vector Goldstone boson, corresponding to the spontaneous violation
of the actual $SO(1,3)\otimes G$ symmetry of the Lagrangian $\mathrm{L}$, $%
D-1$ vector pseudo-Goldstone bosons (PGB) are also produced\footnote{%
Note that in total there appear $4D-1$ pseudo-Goldstone modes, complying
with the number of broken generators of $SO(D,3D)$, both for time-like and
space-like SLIV. From these $4D-1$ pseudo-Goldstone modes, $3D$ modes
correspond to the $D$ three-component vector states as will be shown below,
while the remaining $D-1$ modes are scalar states which will be excluded
from the theory.} due to the breaking of the accidental $SO(D,3D)$ symmetry
of the constraint (\ref{const2}). In contrast to the familiar scalar PGB
case \cite{GLA}, the vector PGBs remain strictly massless being protected by
the simultaneously generated non-Abelian gauge invariance (\ref{nab1}).
Together with the above true vector Goldstone boson, they complete the whole
gauge field multiplet of the internal symmetry group $G$.

After the explicit use of this constraint (\ref{const2}), which constitutes
one supplementary condition on the vector field multiplet $\boldsymbol{A}%
_{\mu }^{a}$, one can identify the pure Goldstone field modes $\boldsymbol{a}%
_{\mu }^{a}$ as follows 
\begin{equation}
\text{\ \ }\boldsymbol{A}_{\mu }^{a}=\boldsymbol{a}_{\mu }^{a}+\frac{%
\boldsymbol{n}_{\mu }^{a}}{\boldsymbol{n}^{2}}(M^{2}-\boldsymbol{n}^{2}%
\boldsymbol{a}^{2})^{\frac{1}{2}}\text{ },\text{ \ }\boldsymbol{n}_{\mu }^{a}%
\boldsymbol{a}^{\mu ,a}\text{\ }=0\text{ \ \ \ }(\boldsymbol{a}^{2}\equiv 
\boldsymbol{a}_{\mu }^{a}\boldsymbol{a}^{\mu ,a}).  \label{sup'}
\end{equation}%
There is also an effective \textquotedblleft Higgs" mode ($\boldsymbol{n}%
_{\mu }^{a}/\boldsymbol{n}^{2})(M^{2}-\boldsymbol{n}^{2}\boldsymbol{a}%
^{2})^{1/2}$ given by the SLIV constraint (one takes again the positive sign
for the square root when expanding it in powers of $\boldsymbol{a}^{2}/M^{2}$%
). Note that, apart from the pure vector fields, the general Goldstonic
modes $\boldsymbol{a}_{\mu }^{a}$ contain $D-1$ scalar modes, $\boldsymbol{a}%
_{0}^{a^{\prime }}$ or $\boldsymbol{a}_{3}^{a^{\prime }}$ ($a^{\prime
}=1...D-1$), for the time-like ($\boldsymbol{n}_{\mu }^{a}=n_{0}^{0}g_{\mu
0}\delta ^{a0}$) or space-like ($\boldsymbol{n}_{\mu }^{a}=n_{3}^{0}g_{\mu
3}\delta ^{a0}$) SLIV respectively. They can be eliminated from the theory
if one imposes appropriate supplementary conditions on the $D-1$ $%
\boldsymbol{a}_{\mu }^{a}$ fields which are still free of constraints. Using
their overall orthogonality (\ref{sup'}) to the physical vacuum direction $%
\boldsymbol{n}_{\mu }^{a}$, one can formulate these supplementary conditions
in terms of a general axial gauge for the entire $\boldsymbol{a}_{\mu }^{a}$
multiplet 
\begin{equation}
n\cdot \boldsymbol{a}^{a}\equiv n_{\mu }\boldsymbol{a}^{\mu ,a}=0,\text{ \ }%
a=0,1,...D-1.  \label{sup''}
\end{equation}%
Here $n_{\mu }$ is the unit Lorentz vector, analogous to that introduced in
the Abelian case, which is now oriented in Minkowskian space-time so as to
be parallel to the vacuum matrix\footnote{%
For such a choice the simple identity $\boldsymbol{n}_{\mu }^{\alpha }\equiv 
\frac{n\cdot \boldsymbol{n}^{\alpha }}{n^{2}}n_{\mu }$ holds, showing that
the rectangular vacuum matrix $\boldsymbol{n}_{\mu }^{\alpha }$ has the
factorized \textquotedblleft two-vector" form.} $\boldsymbol{n}_{\mu }^{a}$.
As a result, in addition to the \textquotedblleft Higgs" mode excluded
earlier by the above orthogonality condition (\ref{sup'}), all the other
scalar fields are eliminated. Consequently only the pure vector fields, $%
\boldsymbol{a}_{i}^{a}$ ($i=1,2,3$ ) or $\boldsymbol{a}_{\mu ^{\prime }}^{a}$
($\mu ^{\prime }=0,1,2$), for time-like or space-like SLIV respectively, are
left in the theory. Clearly, the components $\boldsymbol{a}_{i}^{a=0}$ and $%
\boldsymbol{a}_{\mu ^{\prime }}^{a=0}$ correspond to the true Goldstone
boson, for each type of SLIV respectively, while all the others (for $%
a=1...D-1$) are vector PGBs. Substituting the parameterization (\ref{sup'})
with the SLIV constraint (\ref{const2}) into the Lagrangian (\ref{nab1}) and
expanding the square root in powers of $\boldsymbol{a}^{2}/M^{2}$, one is
led to a highly nonlinear theory in terms of the pure Goldstonic modes $%
\boldsymbol{a}_{\mu }^{a}$. The first and higher order terms in $1/M$ in
this expansion of $\mathrm{L}(\boldsymbol{a}_{\mu }^{a}\boldsymbol{,\psi })$
are Lorentz and $CPT$ violating. Remarkably, however, this theory turns out
to be physically equivalent to a conventional Yang-Mills theory. As was
recently shown \cite{jej}, the Lorentz and $CPT$ violating contributions to
physical processes actually completely cancel out among themselves.
Therefore, the SLIV constraint (\ref{const2}) manifests itself as a
noncovariant gauge condition which does not break physical Lorentz
invariance in the theory.

All the above allows one to conclude that the Yang-Mills theories can
naturally be interpreted as emergent theories caused by SLIV, although
physical Lorentz invariance still remains intact due to the simultaneously
generated gauge invariance. These emergent theories are in fact theories
which provide the building blocks for the Standard Model and beyond, whether
they be exact as in quantum chromodynamics or spontaneously broken as in
grand unified theories and non-Abelian family symmetry models \cite{ram,su3}.

\section{Emergent Tensor Field Gravity}

\subsection{Deriving diffeomorphism invariance}

Let us consider an arbitrary relativistically invariant Lagrangian $\mathcal{%
L}(H_{\mu \nu },\phi )$ for one symmetric two-tensor field $H_{\mu \nu }$
and one real scalar field $\phi $ (chosen as the simplest possible matter)
in the theory taken in Minkowski spacetime. As in vector theories we
restrict ourselves to the minimal dimension interactions. In contrast to
vector fields, whose basic interactions contain dimensionless coupling
constants, interactions with coupling constants of inverse mass
dimensionality (and some of higher powers) are essential for symmetric
tensor fields. Otherwise, one has only a free theory for the spin two
components of the tensor field in the presence of matter fields.

We first turn to the imposition of the SLIV constraint 
\begin{equation}
H_{\mu \nu }H^{\mu \nu }=\mathfrak{n}^{2}M^{2}\text{ , \ \ \ \ }\mathfrak{n}%
^{2}\equiv \mathfrak{n}_{\mu \nu }\mathfrak{n}^{\mu \nu }=\pm 1\text{ }
\label{const3a}
\end{equation}%
on the tensor fields $H_{\mu \nu }$ in the Lagrangian $\mathcal{L}$, which
only possesses global Lorentz (and translational) invariance. Following the
procedure used above for the vector field case, we introduce an extended
Lagrangian $\mathcal{L}^{\prime }$ containing a quadratic Lagrange
multiplier term 
\begin{equation}
\mathcal{L}^{\prime }(H_{\mu \nu },\phi ,\mathcal{\lambda })=\mathcal{L}%
(H_{\mu \nu },\phi )-\frac{1}{4}\mathfrak{\lambda }\left( H_{\mu \nu }H^{\mu
\nu }-\mathfrak{n}^{2}M^{2}\right) ^{2}.  \label{lag22}
\end{equation}%
The variation of $\mathcal{L}^{\prime }(H_{\mu \nu },\phi ,\mathcal{\lambda }%
)$ with respect to $H_{\mu \nu }$ gives\footnote{%
Keeping in mind an application to gravity, we could also admit second order
derivatives of the tensor field \ $H_{\mu \nu }$\ in the Lagrangian $%
\mathcal{L}$ so that the Eulerian $(\mathcal{E}_{H})^{\mu \nu }$ would have
the form \ $(\mathcal{E}_{H})^{\mu \nu }=\partial \mathcal{L}/\partial
H_{\mu \nu }-\partial _{\rho }[\partial \mathcal{L}/\partial (\partial
_{\rho }H_{\mu \nu })]+\partial _{\rho }\partial _{\sigma }[\partial 
\mathcal{L}/\partial (\partial _{\rho }\partial _{\sigma }H_{\mu \nu })]$ .}
the tensor field equation of motion 
\begin{equation}
(\mathcal{E}_{H})^{\mu \nu }-\mathfrak{\lambda }H^{\mu \nu }\left( H_{\rho
\sigma }H^{\rho \sigma }-\mathfrak{n}^{2}M^{2}\right) =0.  \label{eqm2}
\end{equation}%
Here the tensor field Eulerian $(\mathcal{E}_{H})^{\mu \nu }$ is determined
by the starting Lagrangian $\mathcal{L}(H_{\mu \nu },\phi )$, while the
Eulerian of the auxiliary field $\mathfrak{\lambda }(x)$ taken on-shell%
\begin{equation}
\mathcal{E}_{\mathfrak{\lambda }}^{\prime }=\partial \mathcal{L}^{\prime
}/\partial \mathfrak{\lambda }=\frac{1}{4}\left( H_{\mu \nu }H^{\mu \nu }-%
\mathfrak{n}^{2}M^{2}\right) ^{2}=0\text{ \ }  \label{on1}
\end{equation}%
gives the constraint (\ref{const3a}). So, as soon as this constraint holds,
one has the simplified equations of motion%
\begin{equation}
(\mathcal{E}_{H})^{\mu \nu }=0\text{ , \ \ }\mathcal{C}(H_{\mu \nu })=H_{\mu
\nu }H^{\mu \nu }-\mathfrak{n}^{2}M^{2}=0.  \label{eqss}
\end{equation}%
However, due to the quadratic form of the Lagrange multiplier term, the
auxiliary field $\mathfrak{\lambda }(x)$ entirely decouples from the tensor
field dynamics rather than acting as a source of energy-momentum density, as
would be the case if we considered instead a linear Lagrange multiplier term.

The tensor field $H_{\mu \nu }$, both massive and massless, contains many
components which are usually eliminated by imposing some supplementary
conditions\footnote{%
Generally speaking, a symmetric two-tensor field $H_{\mu \nu }$ describes
the states with spin $2$ (five components), spin $1$ (three components) and
two spin $0$ states (each is described by one of its components). Among them
spin $1$ must be necessarily excluded as the sign of the energy for spin $1$
is always opposite to that for spin $2$ and $0$.}. In the massive tensor
field case there are five physical spin-$2$ states to be described by $%
H_{\mu \nu }$. Similarly, in the massless tensor field case, although there
are only two physical (transverse) spin states associated with the graviton,
one cannot construct a symmetric two-tensor field $H_{\mu \nu }$ as a linear
combination of creation and annihilation operators for helicity $\pm 2$
states. It is necessary to add three (and $2j-1,$ in general, for a spin $j$
massless field) fictitious states with other helicities \cite{GLB}. So, in
both the massive and massless tensor field cases, at most five components in
the $10$-component tensor field $H_{\mu \nu }$ may be eliminated and still
preserve Lorentz invariance. Once the SLIV constraint (\ref{const3a}) is
imposed, it follows that only four further supplementary conditions are
possible. In section 3.2 we shall actually only impose three further
supplementary conditions, reducing the number of independent components of $%
H_{\mu \nu }$ to 6 as is done in the Hilbert-Lorentz gauge of general
relativity.

We now turn to the question of the consistency of the SLIV constraint with
the equations of motion for a general symmetric tensor field $H_{\mu \nu }$.
For an arbitrary Lagrangian $\mathcal{L}(H_{\mu \nu },\phi )$, the time
development of the fields would not preserve the constraint. So the
parameters in the Lagrangian must be chosen so as to give a relationship
between the the Eulerians for the tensor and matter fields. In addition to
the lowest dimensional Lorentz covariant expressions constructed from the
Eulerians, we also include the next to lowest dimensional Lorentz covariant
expressions in this relationship. This is necessary in order to allow for
gravitational interactions which vanish in the low energy limit. The lowest
dimensional terms include $\partial _{\mu }(\mathcal{E}_{H})^{\mu \nu }$.
Hence the relationship must transform as a Lorentz vector and carry the
Lorentz index $\mu $ 
\begin{equation}
\mathcal{F}^{\mu }(\mathcal{C}=0;\text{ \ }\mathcal{E}_{H},\mathcal{E}_{\phi
},...)=0\text{ \ \ \ }(\mu =0,1,2,3).  \label{fmu}
\end{equation}%
It therefore takes the following form 
\begin{equation}
\partial _{\mu }(\mathcal{E}_{H})^{\mu \nu }=P_{\alpha \beta }^{\nu }(%
\mathcal{E}_{H})^{\alpha \beta }+Q^{\nu }\mathcal{E}_{\phi }.  \label{idd}
\end{equation}%
Here $P_{\alpha \beta }^{\nu }$ and $Q^{\nu }$ are operators which take the
following general form 
\begin{eqnarray}
P_{\alpha \beta }^{\nu } &=&p_{0}\eta _{\alpha \beta }\partial ^{\nu
}+p_{1}\eta ^{\nu \rho }[H_{\alpha \rho }\partial _{\beta }+H_{\rho \beta
}\partial _{\alpha }+  \label{pp} \\
&&+a(\partial _{\beta }H_{\alpha \rho }+\partial _{\alpha }H_{\rho \beta
})+b\partial _{\rho }H_{\alpha \beta }+cH_{\alpha \beta }\partial _{\rho })]%
\text{ , \ }  \notag \\
Q^{\nu } &=&q_{0}\partial ^{\nu }+q_{1}\eta ^{\nu \rho }(\partial _{\rho
}\phi +d\phi \partial _{\rho }).\text{\ }  \label{ppp}
\end{eqnarray}%
The constants $p_{0}$ and $q_{0}$ are dimensionless and associated with
dimension 4 terms in the relationship, while $p_{1}$ and $q_{1}$ have an
inverse mass dimension and are associated with dimension 5 terms in the
relationship\footnote{%
We note that the double divergence $\partial _{\mu }\partial _{\nu }(%
\mathcal{E}_{H})^{\mu \nu }$ does not appear in (\ref{fmu}, \ref{idd}),
since it would require a term of dimension 6 or higher in order to transform
as a vector.}. In addition $a$, $b$, $c$ and $d$ are as yet undetermined
dimensionless constants. According to Noether's second theorem \cite{noeth},
the identity (\ref{idd}) implies the invariance of the corresponding action 
\begin{equation}
I=\int \mathcal{L}(H_{\mu \nu },\phi )d^{4}x  \label{i}
\end{equation}%
under local transformations of the tensor and scalar fields having the
infinitesimal form 
\begin{eqnarray}
\delta H_{\mu \nu } &=&\partial _{\mu }\xi _{\nu }+\partial _{\nu }\xi _{\mu
}+p_{0}\eta _{\mu \nu }\partial _{\rho }\xi ^{\rho }+p_{1}[\partial _{\mu
}\xi ^{\rho }H_{\rho \nu }+\partial _{\nu }\xi ^{\rho }H_{\mu \rho }+
\label{tr} \\
&&+a(\xi ^{\rho }\partial _{\mu }H_{\rho \nu }+\xi ^{\rho }\partial _{\nu
}H_{\mu \rho })+b\xi ^{\rho }\partial _{\rho }H_{\mu \nu }+c\partial _{\rho
}\xi ^{\rho }H_{\mu \nu }],  \notag \\
\delta \phi  &=&q_{0}\partial _{\rho }\xi ^{\rho }+q_{1}(\xi ^{\rho
}\partial _{\rho }\phi +d\partial _{\rho }\xi ^{\rho }\phi ).  \label{sc}
\end{eqnarray}%
Here $\xi ^{\mu }(x)$ is an arbitrary 4-vector parameter function, only
being required to conform with the nonlinear constraint (\ref{const3}).
These field transformations are treated by themselves as fixed coordinate
system transformations\footnote{%
We shall refer to such transformations as fixed point transformations.},
changing only the functional forms of the fields. One should remember that
we started from a fundamentally flat Minkowski spacetime with only one set
of coordinates (modulo global Lorentz transformations). However it actually
turns out that these field transformations correspond in the end to
reparameterization transformations. Thus it becomes natural to think of
using a modified set of coordinates, deviating from the original fundamental
coordinate system $x^{\mu }$ by $\delta x^{\mu }=x^{\prime \,\mu }-x^{\mu
}\propto \xi ^{\mu }$. In going from the $x^{\mu }$ to the $x^{\prime \,\mu }
$ coordinate system, there are supposed to be infinitesimal coordinate
variations $\delta x^{\mu }$ under which the action $I$ is also left
invariant. The form of these variations will be established later.

In order to avoid generating too many symmetry transformations, which would
only be consistent with a trivial Lagrangian (i.e. $\mathcal{L}=const$), we
further require that the general transformations (\ref{tr}, \ref{sc})
constitute a group. This means that they have to satisfy the Lie bracket
operations 
\begin{eqnarray}
(\delta _{1}\delta _{2}-\delta _{2}\delta _{1})H_{\mu \nu } &=&\delta
_{br}H_{\mu \nu } ,  \label{br1} \\
(\delta _{1}\delta _{2}-\delta _{2}\delta _{1})\phi &=&\delta _{br}\phi. 
\notag
\end{eqnarray}
Here the 4-vector parameter function $\xi _{br}^{\mu }$ related to the Lie
bracket transformation $\delta _{br}$ is supposed to be constructed from the
parameter functions $\xi _{1}^{\mu }$ and $\xi _{2}^{\mu }$, which determine
the single transformations $\delta _{1}$ and $\delta _{2}$ in (\ref{tr}) and
(\ref{sc}). As in the vector field case, this requirement that the Lie
algebra of transformations should close puts strong restrictions on the
values of the constants appearing in (\ref{tr}) and (\ref{sc}). Actually,
after a straightforward calculation similar to that given for the
non-Abelian symmetry case in section 2.2, one finds that the Lie bracket
relations (\ref{br1}) are only satisfied for the following values of the
constants in the field variations $\delta H_{\mu \nu }$ and $\delta \phi $: 
\begin{equation}
\text{ }a=0,\text{ }b=1,\text{ }c=p_{0}\text{ };\text{ \ }q_{0}=0,\text{ }%
q_{1}=p_{1}\text{ }.  \label{rel1}
\end{equation}%
The parameter function $\xi_{br}^{\mu }$ associated with the transformation $%
\delta_{br}$ is given by the expression 
\begin{equation}
\xi _{br}^{\mu }=p_{1}(\xi _{1}^{\rho }\partial _{\rho }\xi _{2}^{\mu }-\xi
_{2}^{\rho }\partial _{\rho }\xi _{1}^{\mu })\text{ .}  \label{br}
\end{equation}%
Remarkably, although the general transformations (\ref{tr}, \ref{sc}) were
only restricted to form a group, the emergent theory turns out to possess a
diffeomorphism invariance provided that the field transformations (\ref{tr}, %
\ref{sc}) are accompanied by an infinitesimal coordinate variation (see
below).

Actually, for the quantity $g_{\mu \nu }$ defined by the equation 
\begin{equation}
g^{p_{0}/2}g_{\mu \nu }=\eta _{\mu \nu }+p_{1}H_{\mu \nu }\text{ , \ \ }%
g\equiv det(g_{\mu \nu }),\text{\ \ }  \label{metr}
\end{equation}%
the transformation (\ref{tr}) may be written in the form\footnote{%
In order to obtain this result, one has first to use the conventional
formulas $\delta g^{p_{0}/2}=$ $(p_{0}/2)g^{p_{0}/2}g^{\alpha \beta} \delta
g_{\alpha \beta }$ and $\partial_{\rho }g^{p_{0}/2}=(p_{0}/2)
g^{p_{0}/2}g^{\alpha \beta }\partial _{\rho }g_{\alpha \beta }$ for the
variation of the determinant $g$ and its derivative respectively, and then
to divide both of sides of the equation by $g^{p_{0}/2}$.}%
\begin{eqnarray}
\delta g_{\mu \nu } &=&p_{1}(\partial _{\mu }\xi ^{\rho }g_{\rho \nu
}+\partial _{\nu }\xi ^{\rho }g_{\mu \rho }+\xi ^{\rho }\partial _{\rho
}g_{\mu \nu })+  \label{mt} \\
&&+p_{0}\left( p_{1}\partial _{\rho }\xi ^{\rho }+\frac{1}{2}p_{1}\xi ^{\rho
}g^{\alpha \beta }\partial _{\rho }g_{\alpha \beta }-\frac{1}{2}g^{\alpha
\beta }\delta g_{\alpha \beta }\right) g_{\mu \nu }\text{ ,}  \notag
\label{deltagmunu}
\end{eqnarray}%
after the above-determined values (\ref{rel1}) of the constants are
substituted in (\ref{tr}). Even in the general case with a non-vanishing
value of the constant $p_{0}$, the explicit solution to this equation for $%
\delta g_{\mu \nu }$ is still given just by the terms independent of $p_0$
in (\ref{deltagmunu}) 
\begin{equation}
\delta g_{\mu \nu }=p_{1}(\partial _{\mu }\xi ^{\rho }g_{\rho \nu }+\partial
_{\nu }\xi ^{\rho }g_{\mu \rho }+\xi ^{\rho }\partial _{\rho }g_{\mu \nu }),
\label{mt1}
\end{equation}%
provided that the contravariant tensor $g^{\mu \nu }$ is properly defined,
that is $g_{\alpha \beta }g^{\beta \gamma }=\delta _{\alpha }^{\gamma }.$ In
fact one can readily verify that 
\begin{equation}
p_{1}\partial _{\rho }\xi ^{\rho }+\frac{1}{2}p_{1}\xi ^{\rho }g^{\alpha
\beta }\partial _{\rho }g_{\alpha \beta }-\frac{1}{2}g^{\alpha \beta }\delta
g_{\alpha \beta }=0\text{ ,}  \label{can}
\end{equation}%
when the expression (\ref{mt1}) is substituted for $\delta g_{\alpha \beta}$
in (\ref{can}). So, one can see that $g_{\mu \nu }$ transforms as the metric
tensor in Riemannian geometry with general coordinate transformations taken
in the form 
\begin{equation}
\delta x^{\mu }=-p_{1}\xi ^{\mu }(x)\text{ .}  \label{co}
\end{equation}%
The constant $p_{1}$ may then be absorbed into the transformation 4-vector
parameter function $\xi ^{\mu }$. Indeed, for this form of the coordinate
variation, the metric changes to 
\begin{equation}
g^{\prime \mu \nu }=\frac{\partial x^{\prime \mu }}{\partial x^{\rho }}\frac{%
\partial x^{\prime \nu }}{\partial x^{\sigma }}g^{\rho \sigma }  \label{met}
\end{equation}%
Plugging in $g^{\mu \nu }=\eta ^{\mu \nu }-p_{1}H^{\mu \nu }+\cdot \cdot
\cdot $ and using $(\partial x^{\prime \mu }/\partial x^{\rho })=\delta
_{\rho }^{\mu }-p_{1}\partial _{\rho }\xi ^{\mu }$, one finds in the weak
field limit (neglecting the terms containing $p_{1}H^{\mu \nu }$ and $\xi
^{\mu }$ altogether and properly lowering the indices with $\eta _{\mu \nu }$
to this order) the reduced transformation law for the tensor field $H_{\mu
\nu }$%
\begin{equation}
\delta H_{\mu \nu }=\partial _{\mu }\xi _{\nu }+\partial _{\nu }\xi _{\mu }%
\text{ .}  \label{tr3}
\end{equation}%
This result conforms with the general equation (\ref{mt1}) taken in the same
limit.

As to the scalar field $\phi (x),$ we can also simplify its transformation
law (\ref{sc}, \ref{rel1}) if we replace it by $\phi ^{\prime }=g^{-d/2}\phi 
$%
\begin{equation}
\delta \phi ^{\prime }=-d/2g^{-d/2}(g^{\alpha \beta }\delta g_{\alpha \beta
})\phi +g^{-d/2}p_{1}(\xi ^{\rho }\partial _{\rho }\phi +d\partial _{\rho
}\xi ^{\rho }\phi )=p_{1}\xi ^{\rho }\partial _{\rho }\phi ^{\prime },
\label{phi1}
\end{equation}%
where we have again used equation (\ref{can}). Therefore, the
transformations for the redefined field $\phi ^{\prime }$ (the prime will be
omitted henceforth) amount to pure local translations.

So we have shown that, in the tensor field case, the imposition of the SLIV
constraint (\ref{const3}) promotes the starting global Poincare symmetry to
the local diff invariance. This SLIV induced gauge symmetry now completely
determines the Lagrangian $\mathcal{L}(H_{\mu \nu },\phi )$ appearing in the
invariant action $I$ (\ref{i}). Actually, as is well-known \cite{kibble}, if
one requires the action integral defined over any arbitrary region to be
invariant (that is, $\delta I=0$) under a total variation, including the
variations of the fields (\ref{tr3}, \ref{phi1}) and of the coordinates (\ref%
{co}), one must have 
\begin{equation}
\delta \mathcal{L}+\partial _{\mu }(\delta x^{\mu }\mathcal{L})=0.
\label{sd}
\end{equation}%
This implies that the Lagrangian $\mathcal{L}(H_{\mu \nu },\phi )$ should
transform like a scalar density rather than being invariant as it usually is
in the internal symmetry case considered in section 2. Now the explicit form
of the Lagrangian $\mathcal{L}(H_{\mu \nu },\phi )$ satisfying the condition
(\ref{sd}), which could be referred to as the action-invariant Lagrangian,
is readily deduced. Indeed, in the weak field approximation, this is the
well-known linearized gravity Lagrangian 
\begin{equation}
\mathcal{L}(H_{\mu \nu },\phi )=\mathcal{L}(H)+\mathcal{L}(\phi )+\mathcal{L}%
_{int}.  \label{tl}
\end{equation}%
It consists of the $H$ field kinetic term of the form%
\begin{equation}
\mathcal{L}(H)=\frac{1}{2}\partial _{\lambda }H^{\mu \nu }\partial ^{\lambda
}H_{\mu \nu }-\frac{1}{2}\partial _{\lambda }H_{tr}\partial ^{\lambda
}H_{tr}-\partial _{\lambda }H^{\lambda \nu }\partial ^{\mu }H_{\mu \nu
}+\partial ^{\nu }H_{tr}\partial ^{\mu }H_{\mu \nu }\text{ ,}  \label{fp}
\end{equation}%
($H_{tr}$ stands for the trace of $H_{\mu \nu },$ $H_{tr}=\eta ^{\mu \nu
}H_{\mu \nu }$) together with the scalar field free Lagrangian part and its
interaction term 
\begin{equation}
\mathcal{L}(\phi )=\frac{1}{2}\left( \partial _{\rho }\phi \partial ^{\rho
}\phi -m^{2}\phi ^{2}\right) \text{ },\text{ \ \ \ \ }\mathcal{L}_{int}=-%
\frac{1}{2M_{P}}H_{\mu \nu }T^{\mu \nu }(\phi )\text{ . \ \ \ \ \ }
\label{fh}
\end{equation}%
Here $T^{\mu \nu }(\phi )$ is the conventional energy-momentum tensor for a
scalar field 
\begin{equation}
T^{\mu \nu }(\phi )=\partial ^{\mu }\phi \partial ^{\nu }\phi -\eta ^{\mu
\nu }\mathcal{L}(\phi )  \label{tt}
\end{equation}%
and the proportionality coefficient $p_{1}$ in the metric (\ref{metr}) is
chosen to be just the inverse Planck mass, $p_{1}=1/M_{P}$. It is clear
that, in contrast to the tensor free field terms given above by $\mathcal{L}%
(H)$, the scalar free field part $\mathcal{L}(\phi )$ and its interaction
term $\mathcal{L}_{int}$ (\ref{fh}) are only approximately action-invariant
under the diff transformations (\ref{tr3}, \ref{phi1}). This only works in
the weak field limit, treating $\partial_{\mu }\xi_{\nu}$ as of the same
order as $H_{\mu \nu }$.

We expect that the reparameterization symmetry will come out to all orders
in $1/M_P$, because the full reparameterization symmetry is needed to ensure
that the equations of motion are free to match with the constraint at all
times. In order to determine the complete theory, one should consider the
full variation of the Lagrangian $\mathcal{L}$ as a function of the metric $%
g_{\mu \nu }$ and its derivatives (including the second order ones) and
solve a general identity of the type 
\begin{equation}
\delta \mathcal{L}(g_{\mu \nu },g_{\mu \nu ,\lambda },g_{\mu \nu ,\lambda
\rho };\phi ,\phi _{,\lambda })=\partial _{\mu }X^{\mu }.  \label{iden}
\end{equation}
Here subscripts after commas denote derivatives and $X^{\mu }$ is an unknown
vector function. The latter must be constructed from the fields and local
transformation parameters $\xi ^{\mu }(x)$, taking into account the
requirement of compatibility with the invariance of $\mathcal{L}$ under
Lorentz transformations and translations. Following this procedure \cite%
{kibble,ogi4} for the field variations (\ref{mt1}, \ref{phi1}) conditioned
by the SLIV constraint (\ref{const3}), one can eventually find the total
Lagrangian $\mathcal{L}$. The latter turns out to be properly expressed in
terms of quantities similar to the basic ones in Riemannian geometry (like
the metric, connection, curvature etc.). Actually, this theory successfully
mimics general relativity, which allows us to conclude that the Einstein
equations can really be derived in flat Minkowski spacetime provided that
the Lorentz symmetry is spontaneously broken.

While we will mainly be focused, in what follows, on the linearized gravity
theory case, our discussion can be extended to general relativity as well.

\subsection{\protect\bigskip Graviton as a tensor Goldstone boson}

Let us turn now to the spontaneous Lorentz violation which is caused by the
nonlinear tensor field constraint (\ref{const3}). This constraint can be
written in the more explicit form 
\begin{equation}
H_{\mu \nu }^{2}=H_{00}^{2}+H_{i=j}^{2}+(\sqrt{2}H_{i\neq j})^{2}-(\sqrt{2}%
H_{0i})^{2}=\mathfrak{n}^{2}M^{2}=\pm \text{ }M^{2}\text{ \ \ \ \ \ \ \ }
\label{c4}
\end{equation}%
(where the summing on indices $(i,j=1,2,3)$ is imposed) and means in essence
that the tensor field $H_{\mu \nu }$ develops the vev configuration 
\begin{equation}
<H_{\mu \nu }(x)>\text{ }=\mathfrak{n}_{\mu \nu }M  \label{v}
\end{equation}%
determined by the matrix $\mathfrak{n}_{\mu \nu }$. The initial Lorentz
symmetry $SO(1,3)$ of the Lagrangian $\mathcal{L}(H_{\mu \nu },\phi )$ given
in (\ref{tl}) then formally breaks down at a scale $M$ to one of its
subgroups. We assume for simplicity a ``minimal" vacuum configuration in the 
$SO(1,3)$ space with the vevs (\ref{v}) developed on only one of the $H_{\mu
\nu }$ components. If so, there are in fact the following three possibilities%
\begin{eqnarray}
(a)\text{ \ \ \ \ }\mathfrak{n}_{00} &\neq &0\text{ , \ \ }%
SO(1,3)\rightarrow SO(3)  \notag \\
(b)\text{ \ \ \ }\mathfrak{n}_{i=j} &\neq &0\text{ , \ \ }SO(1,3)\rightarrow
SO(1,2)  \label{ns} \\
(c)\text{ \ \ \ }\mathfrak{n}_{i\neq j} &\neq &0\text{ , \ \ }%
SO(1,3)\rightarrow SO(1,1)  \notag
\end{eqnarray}%
for the positive sign in (\ref{c4}), and 
\begin{equation}
(d)\text{ \ \ \ }\mathfrak{n}_{0i}\neq 0\text{ , \ \ }SO(1,3)\rightarrow
SO(2)  \label{nss}
\end{equation}%
for the negative sign. These breaking channels can be readily derived, by
counting how many different eigenvalues the matrix $\mathfrak{n}_{\mu \nu }$
has for each particular case ($a$-$d$). Accordingly, there are only three
Goldstone modes in the cases ($a,b$) and five modes in the cases ($c$-$d$).
In order to associate at least one of the two transverse polarization states
of the physical graviton with these modes, one could have any of the
above-mentioned SLIV channels except for the case ($a$). Indeed, it is
impossible for the graviton to have all vanishing spatial components, as
happens for the Goldstone modes in the case ($a$). Therefore, no linear
combination of the three Goldstone modes in case ($a$) could behave like the
physical graviton (see \cite{car} for a more detailed consideration). In
addition to the minimal vev configuration, there are many other
possibilities. A particular case of interest is that of the traceless vev
tensor $\mathfrak{n}_{\mu \nu }$%
\begin{equation}
\text{\ \ }\mathfrak{n}_{\mu \nu }\eta ^{\mu \nu }=0,  \label{tll}
\end{equation}%
in terms of which the Goldstonic gravity Lagrangian acquires an especially
simple form (see below). It is clear that the vev in this case can be
developed on several $H_{\mu \nu }$ components simultaneously, which in
general may lead to total Lorentz violation with all six Goldstone modes
generated. For simplicity we will use this form of vacuum configuration in
what follows, while our arguments can be applied to any type of vev tensor $%
\mathfrak{n}_{\mu \nu }.$

In this connection the question naturally arises of the other components of
the symmetric two-index tensor $H_{\mu \nu },$ in addition to the pure
Goldstone modes. They turn out to be pseudo-Goldstone modes (PGMs) in the
theory. Indeed, although we only propose Lorentz invariance of the
Lagrangian $\mathcal{L}(H_{\mu \nu },\phi )$, the SLIV constraint (\ref%
{const3}) formally possesses the much higher accidental symmetry $SO(7,3)$
of the constrained bilinear form (\ref{c4}), when the $H_{\mu \nu }$
components are considered as the ``vector" components under $SO(7,3)$. This
symmetry is in fact spontaneously broken side by side with Lorentz symmetry
at the scale $M$. Assuming again a minimal vacuum configuration in the $%
SO(7,3)$ space with the vev (\ref{v}) developed on only one of the $H_{\mu
\nu }$ components, we have either time-like ($SO(7,3)$ $\rightarrow SO(6,3)$%
) or space-like ($SO(7,3)$ $\rightarrow SO(7,2)$) violations of the
accidental symmetry depending on the sign of $\mathfrak{n}^{2}=\pm 1$ in (%
\ref{c4}). According to the number of broken $SO(7,3)$ generators, just nine
massless NG modes appear in both cases. Together with an effective Higgs
component, on which the vev is developed, they complete the whole
ten-component symmetric tensor field $H_{\mu \nu }$ of our Lorentz group.
Some of them are true Goldstone modes of the spontaneous Lorentz violation.
The others are PGMs since the accidental $SO(7,3)$ is not shared by the
whole Lagrangian $\mathcal{L}(H_{\mu \nu },\phi )$ given in (\ref{tl}).
Notably, in contrast to the scalar PGM case \cite{GLA} and similarly to the
vector PGMs, they remain strictly massless being protected by the
simultaneously generated diff invariance\footnote{%
For a non-minimal vacuum configuration when vevs are developed on several \ $%
H_{\mu \nu }$\ components, thus leading to a more substantial breaking of
the accidental $SO(7,3)$ symmetry, some extra PGMs are generated. However,
they are not protected by diffeomorphism invariance and acquire masses of
the order of the breaking scale ($M$).}. Owing to the latter invariance,
some of the PGMs and Goldstone modes can be gauged away from the theory, as
usual.

Now, one can rewrite the Lagrangian $\mathcal{L}(H_{\mu \nu },\phi )$ in
terms of the Goldstone modes explicitly using the SLIV constraint (\ref%
{const3}). For this purpose let us take the following handy parameterization
for the tensor field $H_{\mu \nu }$ in the Lagrangian $\mathcal{L}(H_{\mu
\nu },\phi )$: 
\begin{equation}
H_{\mu \nu }=h_{\mu \nu }+\frac{\mathfrak{n}_{\mu \nu }}{\mathfrak{n}^{2}}(%
\mathfrak{n}\cdot H)\qquad (\mathfrak{n}\cdot H\equiv \mathfrak{n}_{\mu \nu
}H^{\mu \nu }),  \label{par}
\end{equation}%
where $h_{\mu \nu }$ corresponds to the pure Goldstonic modes\footnote{%
It should be particularly emphasized that the modes collected in $h_{\mu\nu
} $ are in fact the Goldstone modes of the broken accidental $SO(7,3)$
symmetry of the constraint (\ref{const3}) thus containing the Lorentz
Goldstone modes and PGMs altogether.} satisfying 
\begin{equation}
\text{\ }\mathfrak{n}\cdot h=0\text{\ }\qquad (\mathfrak{n}\cdot h\equiv 
\mathfrak{n}_{\mu \nu }h^{\mu \nu }).  \label{sup}
\end{equation}%
There is also an effective \textquotedblleft Higgs" mode (or the $H_{\mu \nu
}$ component in the vacuum direction) is given by the scalar product $%
\mathfrak{n}\cdot H$. Substituting this parameterization (\ref{par}) into
the tensor field constraint (\ref{const3}), one obtains the following
equation for $\mathfrak{n}\cdot H$: 
\begin{equation}
\text{\ }\mathfrak{n}\cdot H\text{\ }=(M^{2}-\mathfrak{n}^{2}h^{2})^{\frac{1%
}{2}}=M-\frac{\mathfrak{n}^{2}h^{2}}{2M}+O(1/M^{2})  \label{constr1}
\end{equation}%
taking, for definiteness, the positive sign for the square root and
expanding it in powers of $h^{2}/M^{2}$, $h^{2}\equiv h_{\mu \nu }h^{\mu \nu
}$. Putting then the parameterization (\ref{par}) with the SLIV constraint (%
\ref{constr1}) into the Lagrangian $\mathcal{L}(H_{\mu \nu },\phi )$ given
in (\ref{fp}, \ref{fh}), one obtains the Goldstonic tensor field gravity
Lagrangian $\mathcal{L}(h_{\mu \nu },\phi )$ containing an infinite series
in powers of the $h_{\mu \nu }$ modes. For the traceless vev tensor $%
\mathfrak{n}_{\mu \nu }$ (\ref{tll}) it takes, without loss of generality,
the especially simple form 
\begin{eqnarray}
\mathcal{L}(h_{\mu \nu },\phi ) &=&\frac{1}{2}\partial _{\lambda }h^{\mu \nu
}\partial ^{\lambda }h_{\mu \nu }-\frac{1}{2}\partial _{\lambda
}h_{tr}\partial ^{\lambda }h_{tr}-\partial _{\lambda }h^{\lambda \nu
}\partial ^{\mu }h_{\mu \nu }+\partial ^{\nu }h_{tr}\partial ^{\mu }h_{\mu
\nu }+  \label{gl} \\
&&+\frac{1}{2M}h^{2}\left[ -2\mathfrak{n}^{\mu \lambda }\partial _{\lambda
}\partial ^{\nu }h_{\mu \nu }+\mathfrak{n}^{2}(\mathfrak{n}\partial \partial
)h_{tr}\right] +\frac{1}{8M^{2}}h^{2}\left[ -\mathfrak{n}^{2}\partial
^{2}+2(\partial \mathfrak{nn}\partial )\right] h^{2}  \notag \\
&&+\mathcal{L}(\phi )-\frac{M}{2M_{P}}\mathfrak{n}^{2}\left[ \mathfrak{n}%
_{\mu \nu }\partial ^{\mu }\phi \partial ^{\nu }\phi \right] -\frac{1}{2M_{P}%
}h_{\mu \nu }T^{\mu \nu }-\frac{1}{4MM_{P}}h^{2}\left[ -\mathfrak{n}_{\mu
\nu }\partial ^{\mu }\phi \partial ^{\nu }\phi \right]  \notag
\end{eqnarray}%
written in the $O(h^{2}/M^{2})$ approximation. In addition to the
conventional graviton bilinear kinetic terms, the Lagrangian contains three-
and four-linear interaction terms in powers of $h_{\mu \nu }$. Some of the
notations used are collected below: 
\begin{eqnarray}
h^{2} &\equiv &h_{\mu \nu }h^{\mu \nu }\text{ , \ \ }h_{tr}\equiv \eta ^{\mu
\nu }h_{\mu \nu }\text{ , \ }  \label{n} \\
\mathfrak{n}\partial \partial &\equiv &\mathfrak{n}_{\mu \nu }\partial ^{\mu
}\partial ^{\nu }\text{\ , \ \ }\partial \mathfrak{nn}\partial \equiv
\partial ^{\mu }\mathfrak{n}_{\mu \nu }\mathfrak{n}^{\nu \lambda }\partial
_{\lambda }\text{ .\ \ \ }  \notag
\end{eqnarray}

The bilinear scalar field term%
\begin{equation}
-\frac{M}{2M_{P}}\mathfrak{n}^{2}\left[ \mathfrak{n}_{\mu \nu }\partial
^{\mu }\phi \partial ^{\nu }\phi \right]  \label{t}
\end{equation}%
in the third line in the Lagrangian (\ref{gl}) merits special notice. This
term arises from the interaction Lagrangian $\mathfrak{L}_{int}$ (\ref{fh})
after application of the tracelessness condition (\ref{tll}) for the vev
tensor $\mathfrak{n}_{\mu \nu }$. It could significantly affect the
dispersion relation for the scalar field $\phi $ (and any other sort of
matter as well), thus leading to an unacceptably large Lorentz violation if
the SLIV scale $M$ were comparable with the Planck mass $M_{P}.$ However,
this term can be gauged away by an appropriate choice of the gauge parameter
function $\xi ^{\mu }(x)$ in the transformations (\ref{tr3}, \ref{phi1}) of
the tensor and scalar fields\footnote{%
Actually, in the Lagrangian $\mathcal{L}(H_{\mu \nu },\phi )$ satisfying the
action invariance condition (\ref{sd}), the vacuum shift of the tensor field 
$H_{\mu \nu }=h_{\mu \nu }+\frac{\mathfrak{n}_{\mu \nu }}{\mathfrak{n}^{2}}M$
is in fact a gauge transformation which, for the appropriately chosen
transformation of the scalar field $\phi (x),$ leaves the action $I$ (\ref{i}%
) invariant.}. Technically, one simply transforms the scalar field and its
derivative to a new coordinate system $x^{\mu }\rightarrow $ $x^{\mu }-\xi
^{\mu }$ in the Goldstonic Lagrangian $\mathcal{L}(h_{\mu \nu },\phi )$.
Actually, using the fixed-point variation of $\phi (x)$ given above in (\ref%
{phi1}), with the coefficient $p_{1}$ absorbed into the parameter function $%
\xi ^{\mu }(x)$, and differentiating both sides with respect to $x^{\mu }$
one obtains 
\begin{equation}
\delta (\partial _{\mu }\phi )=\partial _{\mu }(\xi ^{\nu }\partial _{\nu
}\phi ).
\end{equation}%
This gives in turn 
\begin{equation}
\delta _{tot}(\partial _{\mu }\phi )=\delta (\partial _{\mu }\phi )+\delta
x^{\nu }\partial _{\nu }(\partial _{\mu }\phi )= \partial _{\mu }\xi ^{\nu
}\partial _{\nu }\phi  \label{red}
\end{equation}%
for the total variation of the scalar field derivative. The corresponding
total variation of the Goldstonic tensor $h_{\mu \nu }$, caused by the same
transformation to the coordinate system $x^{\mu }-\xi ^{\mu }$, is given in
turn by equations (\ref{tr3}) and (\ref{par}) to be 
\begin{equation}
\delta _{tot}h_{\mu \nu }=(\partial ^{\rho }\xi ^{\sigma }+\partial ^{\sigma
}\xi ^{\rho })\left( \eta _{\rho \mu }\eta _{\sigma \nu }-\frac{\mathfrak{n}%
_{\mu \nu }}{\mathfrak{n}^{2}}\mathfrak{n}_{\rho \sigma }\right) -\xi ^{\rho
}\partial _{\rho }h_{\mu \nu }.  \label{dh}
\end{equation}%
One can now readily see that, with the parameter function $\xi ^{\mu }(x)$
chosen as 
\begin{equation}
\xi ^{\mu }(x)= \frac{M}{2M_{P}}\mathfrak{n}^{2}\mathfrak{n}^{\mu \nu
}x_{\nu }\text{ },
\end{equation}%
the dangerous term (\ref{t}) is precisely cancelled\footnote{%
In the general case, with the vev tensor $\mathfrak{n}_{\mu \nu }$ having a
non-zero trace, this cancellation would also require the redefinition of the
scalar field itself as $\phi \rightarrow \phi (1-\mathfrak{n}_{\mu \nu }\eta
^{\mu \nu }\frac{M}{M_{P}})^{-1/2}$.} by an analogous term stemming from the
scalar field kinetic term in the $\mathfrak{L}(\phi )$ given in (\ref{fh}),
while the total variation of the tensor $h_{\mu \nu }$ reduces to just the
second term in (\ref{dh}). This term is of the natural order $O(\xi h)$,
which can be neglected in the weak field approximation, so that to the
present accuracy the tensor field variation $\delta _{tot}h_{\mu\nu }=0$.
Indeed, since the diff invariance is an approximate symmetry of the
Lagrangian $\mathcal{L}(h_{\mu \nu },\phi )$, the above cancellation will
only be accurate up to the order corresponding to the linearized Lagrangian $%
\mathcal{L}(H_{\mu \nu },\phi )$ we started with in (\ref{tl}). Actually, a
proper extension of the tensor field theory to GR with its exact diff
invariance will ultimately restore the usual form of the dispersion relation
for the scalar (and other matter) fields. Taking this into account, we will
henceforth omit the term (\ref{t}) in $\mathcal{L}(h_{\mu \nu },\phi )$,
thus keeping the ``normal" dispersion relation for the scalar field in what
follows.

Together with the Lagrangian one must also specify the other gauge fixing in
addition to the general Goldstonic \textquotedblleft gauge" \ $\mathfrak{n}%
_{\mu \nu }\cdot h^{\mu \nu }=0$ choice given above (\ref{sup}). The point
is that the spin $1$ states are still left in the theory\footnote{%
These spin $1$ states must necessarily be excluded as the sign of the energy
for spin $1$ is always opposite to that for spin $2$ and $0$} and are
described by some of the components of the new tensor $h_{\mu \nu }$.
Usually, they (and one of the spin $0$ states) are excluded by the
conventional Hilbert-Lorentz condition 
\begin{equation}
\partial ^{\mu }h_{\mu \nu }+q\partial ^{\nu }h_{tr}=0  \label{HL}
\end{equation}%
($q$ is an arbitrary constant giving the standard harmonic gauge condition
for $q=-1/2$). On the other hand, as we have already imposed the constraint (%
\ref{sup}), we cannot use the full Hilbert-Lorentz condition (\ref{HL}),
eliminating four more degrees of freedom in $h_{\mu \nu }.$ Otherwise, one
would have an \textquotedblleft overgauged" theory with a non-propagating
graviton. In fact the simplest set of conditions which conforms with the
Goldstonic condition (\ref{sup}) turns out to be \cite{cjt} 
\begin{equation}
\partial ^{\rho }(\partial _{\mu }h_{\nu \rho }-\partial _{\nu }h_{\mu \rho
})=0  \label{gauge}
\end{equation}%
This set excludes only three degrees of freedom\footnote{%
The solution for the gauge function $\xi _{\mu }(x)$ satisfying the condition%
$\ $(\ref{gauge}) can generally be chosen to be $\xi _{\mu }=$\ $\ \square
^{-1}(\partial ^{\rho }h_{\mu \rho })+\partial _{\mu }\theta $ where $\theta
(x)$ is an arbitrary scalar function, so that only three degrees of freedom
in $h_{\mu \nu }$ are actually eliminated.} in $h_{\mu \nu }$ and it
automatically satisfies the Hilbert-Lorentz spin condition as well. So, with
the Lagrangian (\ref{gl}) and the supplementary conditions\ (\ref{sup}) and (%
\ref{gauge}) lumped together, one eventually comes to a working model for
the Goldstonic tensor field gravity. Generally, from ten components in the
symmetric-two $h_{\mu \nu }$ tensor, four components are excluded by the
supplementary conditions (\ref{sup}) and (\ref{gauge}). For a plane
gravitational wave propagating, say, in the $z$ direction another four
components can also be eliminated. This is due to the fact that the above
supplementary conditions still leave freedom in the choice of a coordinate
system, $x^{\mu }\rightarrow $ $x^{\mu }-\xi ^{\mu }(t-z/c),$ much as takes
place in standard GR. Depending on the form of the vev tensor $\mathfrak{n}%
_{\mu \nu }$, the two remaining transverse modes of the physical graviton
may consist solely of Lorentz Goldstone modes or of Pseudo Goldstone modes
or include both of them.

The theory derived looks essentially nonlinear and contains a variety of
Lorentz (and $CPT$) violating couplings, when expressed in terms of the pure
tensor Goldstone modes. Nonetheless, as was shown in recent calculations 
\cite{cjt}, all the SLIV effects turn out to be strictly cancelled in the
lowest order graviton-graviton scattering processes, due to the exact
diffeomorphism invariance of the pure gravity part in the basic Lagrangian $%
\mathcal{L}$ (\ref{gl}). At the same time, an actual Lorentz violation may
appear in the matter field interaction sector, which only possesses an
approximate diff invariance, through deformed dispersion relations of the
matter fields involved. However, a proper extension of the tensor field
theory to GR with its exact diffeomorphism invariance ultimately restores
the dispersion relations for matter fields and, therefore, the SLIV effects
vanish. So, one could generally argue, the measurable effects of SLIV,
induced by elementary vector or tensor fields, can be related to the
accompanying gauge symmetry rather than to spontaneous Lorentz violation.
The latter appears by itself to be physically unobservable and only results
in a noncovariant gauge choice in an otherwise gauge invariant and Lorentz
invariant theory.

From this standpoint, the only way for physical Lorentz violation to appear
would be if the above local invariance is slightly broken at very small
distances in an explicit, rather than spontaneous, way. This is in fact a
place where the emergent vector and tensor field theories may differ from
conventional QED, Yang-Mills and GR theories. Actually, such a local
symmetry breaking could lead in the former case to deformed dispersion
relations for all the matter fields involved. This effect typically appears
proportional to some power of the ratio $\frac{M}{M_{P}}$ (just as we have
seen above for the scalar field in our model, see (\ref{t})), though being
properly suppressed due to the tiny gauge noninvariance. The higher the SLIV
scale $M$ becomes the larger becomes the actual Lorentz violation which, for
some value of the scale $M$, may become physically observable even at low
energies. Another basic difference between Goldstonic theories with
non-exact gauge invariance and conventional theories is the emergence of a
mass for the graviton and other gauge fields (namely, for the non-Abelian
ones), if they are composed from Pseudo Goldstone modes rather than from
pure Goldstone ones. Indeed, these PGMs are no longer protected by gauge
invariance and may acquire tiny masses. This may lead to a massive gravity
theory, where the graviton mass emerges dynamically, thus avoiding the
notorious discontinuity problem \cite{zvv}. So, while Goldstonic theories
with exact local invariance are physically indistinguishable from
conventional gauge theories, there are some principal differences when this
local symmetry is slightly broken which could eventually allow us to
differentiate between them in an observational way.

One could imagine how such a breaking might occur. As we have learned, only
locally invariant theories provide the needed number of degrees of freedom
for the interacting vector fields once SLIV occurs. Note that a superfluous
restriction on a vector (or any other) field would make it impossible to set
the required initial conditions in the appropriate Cauchy problem and, in
quantum theory, to choose self-consistent equal-time commutation relations 
\cite{ogi3}. One could expect, however, that quantum gravity could in
general hinder the setting of the required initial conditions at extra-small
distances. Eventually this would manifest itself in an explicit violation of
the above local invariance in a theory through some high-order operators
stemming from the quantum gravity energy scale, which could lead to physical
Lorentz violation. If so, one could have some observational evidence in
favor of the emergent theories, just as was claimed at the very beginning
when the SLIV idea was put forward \cite{bjorken}. However, is there really
any strong theoretical reason left for the Lorentz invariance to be
physically broken, if the Goldstonic gauge fields are anyway generated
through the \textquotedblleft safe" nonlinear sigma type SLIV models which
recover conventional Lorentz invariance? We may return to this question
elsewhere.

\section{Conclusion}

An arbitrary local theory of a symmetric two-tensor field $H_{\mu \nu }$ in
Minkowski spacetime was considered, in which the equations of motion are
required to be compatible with a nonlinear length-fixing constraint $H_{\mu
\nu }^{2}=\pm M^{2}$ leading to spontaneous Lorentz invariance violation ($M$
is the proposed scale for SLIV). Allowing the parameters in the Lagrangian
to be adjusted so as to be compatible with this constraint, the theory turns
out to correspond to general relativity (in the weak field approximation).
Also some of the massless tensor Goldstone modes appearing through SLIV are
naturally collected in the physical graviton. The underlying diffeomophism
invariance directly follows from an application, of Noether's second theorem 
\cite{noeth}. In fact we argued for a relation between the Eulerians
(equation of motion expressions), which then by Noether's second theorem
implies the reparameterization symmetry of the Lagrangian. Such a relation (%
\ref{fmu}, \ref{idd}) is needed for consistency, when the constraint $H_{\mu
\nu }^{2}=\pm M^{2}$ is to be upheld at all times. Otherwise the degrees of
freedom of the symmetric two-tensor $H_{\mu \nu }$ would be superfluously
restricted. Actually, this derivation of diffeomorphism symmetry excludes
\textquotedblleft wrong\textquotedblright\ couplings in the tensor field
Lagrangian, which would otherwise distort the final Lorentz symmetry broken
phase with unphysical extra states including ghost-like ones. Note that this
procedure might, in some sense, be inspired by string theory where the
coupling constants are just vacuum expectation values of the dilaton and
moduli fields \cite{string}. So, the adjustment of coupling constants in the
Lagrangian would mean, in essence, a certain choice for the vacuum
configurations of these fields, which are thus correlated with SLIV.

The crucial point in our method of deriving gauge invariance seems to be
that one degree of freedom for the vector or tensor field considered is not
determined from the time development of their own equations of motion but
solely by the relevant constraint (\ref{const}, \ref{const3}, \ref{const2}).
So, in order to avoid a possible inconsistency with an accordingly
diminished number of independent degrees of freedom for the fields involved,
their equations of motion must be generically prearranged to have less
predictive power. Such a reduced predictive power is precisely what is
achieved in gauge theories, where one cannot predict the evolution of
gauge-fixing terms as time develops. The equations of motion in gauge
theories are therefore less predictive by just the number of degrees of
freedom corresponding to the number of gauge parameters, which are actually
functions of spacetime. In order to allow for consistency with constraints
like (\ref{const}, \ref{const3}, \ref{const2}), one at first seems to need
that the number of gauge degrees of freedom should be equal to the number of
such constraints. But, as we have seen, even one constraint introduced as a
length-fixing condition (\ref{const}, \ref{const3}, \ref{const2}) may be
enough for several gauge symmetry generators to emerge. Such a length-fixing
constraint (\ref{const}) applied to the one-vector field case (Section 2.1)
leads to QED with only one gauge degree of freedom given by a gauge function 
$\omega (x)$. However, the analogous constraint (\ref{const2}) in the
non-Abelian case, with the starting global $G$ symmetry (Section 2.2),
requires that $D$ conditions $\boldsymbol{F}_{a}(C=0;\text{ \ }\mathrm{E}_{%
\boldsymbol{A}},\mathrm{E}_{\boldsymbol{\psi }},...)=0$ have to be
simultaneously fulfilled. This eventually leads to a gauge invariant
Yang-Mills theory with $D$ gauge degrees of freedom given by the set of
parameter functions $\omega ^{a}(x).$ Similarly in the tensor field case
(Section 3.1), the length-fixing constraint (\ref{const2}) requires that
just four equations $\mathcal{F}^{\mu } (\mathcal{C}=0;\text{ \ }\mathcal{E}%
_{H},\mathcal{E}_{\phi},...)=0$ should be arranged to be automatically
satisfied. This leads to the diffeomorphism invariance (\ref{tr}) with the
transformation 4-vector parameter function $\xi ^{\mu }(x).$

The appearance of gauge symmetries in our approach hinges strongly upon the
imposition of a constraint. This can be done in either of the two following
ways: (1) the constraint is imposed by hand prior to varying of the action
or (2) the constraint is imposed by introducing a special quadratic Lagrange
multiplier term, for which the Lagrange multiplier field is decoupled from
the equations of motion and is thereby unable to ensure their consistency
with the constraint. In both cases it is not possible to have consistency
between the equations of motion and the constraint, unless the parameters in
the Lagrangian are adjusted to allow for more freedom in the time
development. This typically means that the Lagrangian should possess a
generic, SLIV enforced, gauge invariance. As a result, all these vector and
tensor field theories do not lead to any physical Lorentz violation and are
in fact indistinguishable from conventional QED, Yang-Mills theories and
general relativity\footnote{%
Nonetheless, imposing nonlinear constraints in the emergent theories raises
the question of unitarity and stability in them. Indeed, while the gauge
invariant form for the vector (tensor) field kinetic terms in them prevents
propagation of their longitudinal modes as the ghost modes, these nonlinear
gauge conditions could cause them unless the phase space in these theories
are properly restricted so as to have ghost-free models with positive
Hamiltonians. Particularly, it was shown [30] that by restricting the phase
space to the vector field solutions with initial values obeying Gauss's law,
the equivalence of Nambu's nonlinear QED model with an ordinary ghost-free
QED is restored. At the same time, if these constraints are introduced, as
in our case, through the quadratic Lagrange multiplier potentials (7, 17,
39) then a Hamiltonian appears positive over the full phase space [30].
Though these results have been still established for Abelian case only, one
could expect that similar arguments are applicable to all gauge theories
considered.}. However, there might appear some principal distinctions if
these emergent local symmetries were slightly broken at very small distances
controlled by quantum gravity in an explicit, rather than spontaneous, way
that could eventually allow one to differentiate between emergent and
conventional gauge theories observationally.

\section*{Acknowledgments}

One of us (J.L.C.) appreciates the warm hospitality shown to him during his
visit to the Division of Elementary Particle Physics, Department of Physics,
University of Helsinki where part of this work was carried out. We would
like to thank Masud Chaichian, Oleg Kancheli, Archil Kobakhidze, Rabi
Mohapatra and Giovanni Venturi for useful discussions and comments.
Financial support from Georgian National Science Foundation (grant N
07\_462\_4-270) is gratefully acknowledged by J.L.C. Also C.D.F. would like
to acknowledge support from STFC in UK.

\end{document}